\documentclass[11pt,english]{article}
\pdfoutput=1

\usepackage{setspace}
\usepackage{babel}
\usepackage{graphicx}
\usepackage{amsmath}
\usepackage{amssymb}
\usepackage{cite}
\usepackage{float}
\usepackage[colorlinks=true,urlcolor=blue,anchorcolor=blue,citecolor=blue,filecolor=blue,linkcolor=blue,menucolor=blue,pagecolor=blue,linktocpage=true]{hyperref}
\usepackage{array}

\newcommand{\dho}{\partial}

\newcommand{\cc}{\mathrm{c.c.}}
\newcommand{\ed}{\,.}
\newcommand{\ec}{\,,}
\newcommand{\ecq}{\ec\quad}

\newcommand{\bZ}{\ensuremath{\mathbb{Z}}}

\newcommand{\cO}{\ensuremath{\mathcal{O}}}

\newcommand{\Li}{\ensuremath{\mathrm{Li}}}

\newcommand{\tily}{\ensuremath{\tilde{y}}}

\let\pa=\partial
\def\be{\begin{equation}}
\def\ee{\end{equation}}
\def\beq{\begin{equation}}
\def\eeq{\end{equation}}
\def\ba{\begin{array}}
\def\ea{\end{array}}

\newcommand{\rchi}{{\mathpalette\irchi\relax}}
\newcommand{\irchi}[2]{\raisebox{\depth}{$#1\chi$}}

\newcommand{\bos}{b}
\newcommand{\barx}{\bar{x}}
\newcommand{\bary}{\bar{y}}
\newcommand{\barL}{\bar{\Lambda}}
\newcommand{\barmu}{\bar{\mu}}

\newcommand{\bra}[1]{\left\langle #1 \right|}
\newcommand{\ket}[1]{\left| #1 \right\rangle}

\DeclareMathOperator{\trace}{Tr}
\DeclareMathOperator{\sign}{sign}

\begin{document}

\begin{flushright} 
\today\\

SU-ITP-16/09

\end{flushright} 

\vspace{0.1cm}

\begin{center}
  {\LARGE
  Transport in Chern-Simons-Matter Theories
}
\end{center}
\vspace{0.1cm}
\vspace{0.1cm}
\begin{center}
	
      Guy G{\sc ur-Ari}\footnote
      {E-mail: guyga@stanford.edu},  
      Sean H{\sc artnoll}\footnote
      {E-mail: hartnoll@stanford.edu}
and
      Raghu M{\sc ahajan}\footnote
      {E-mail: rm89@stanford.edu}
          
\vspace{0.5cm}
	
{\it Stanford Institute for Theoretical Physics,\\
Stanford University, Stanford, CA 94305, USA}

\end{center}

\onehalfspacing

\vspace{1.5cm}

\begin{center}
  {\bf Abstract}
\end{center}

The frequency-dependent longitudinal and Hall conductivities --- $\sigma_{xx}$ and $\sigma_{xy}$ --- are dimensionless functions of $\omega/T$ in 2+1 dimensional CFTs at nonzero temperature. These functions characterize the spectrum of charged excitations of the theory and are basic experimental observables. We compute these conductivities for large $N$ Chern-Simons theory with fermion matter. The computation is exact in the 't Hooft coupling $\lambda$ at $N = \infty$. We describe various physical features of the conductivity, including an explicit relation between the weight of the delta function at $\omega = 0$ in $\sigma_{xx}$ and the existence of infinitely many higher spin conserved currents in the theory. We also compute the conductivities perturbatively in Chern-Simons theory with scalar matter and show that the resulting functions of $\omega/T$ agree with the strong coupling fermionic result. This provides a new test of the conjectured 3d bosonization duality. In matching the Hall conductivities we resolve an outstanding puzzle by carefully treating an extra anomaly that arises in the regularization scheme used.

\vspace{0.1cm}

\newpage

\tableofcontents

\newpage

\section{Background}

The matrix of electrical conductivities
\be
\sigma_{ab}(\omega) = \lim_{z \to \omega + i 0^+} \frac{1}{i z} \left[ G^R_{J_a J_b}(z) - \rchi_{ab} \right]
 \,,\label{eq:first}
\ee
is a basic physical observable of systems with a global $U(1)$ symmetry. Here $G^R$ is the retarded Green's function of the spatial components of the current operator $\vec J$, and $\rchi$ is the Euclidean Green's function evaluated at the Matsubara frequency $\omega_n=0$, 
i.e. $\rchi_{ab} = \langle J_a J_b\rangle(\omega_n=0)$ \cite{Luttinger:1964zz, forster}. Throughout, the external spatial momentum $\vec k = 0$. At a quantum critical point described by a 2+1 dimensional Conformal Field Theory (CFT), scaling symmetry and the protected dimension of the current operator imply that $\sigma_{ab}$ is a dimensionless function of frequency over temperature \cite{subirbook}
\be
\sigma_{ab}(\omega,T) = \sigma_{ab}\left(\frac{\omega}{T} \right) \,. \label{eq:acac}
\ee
In particular, this means \cite{PhysRevB.56.8714, PhysRevB.57.7157} that the d.c.~conductivity ($\omega \to 0$) is generically a distinct quantity to the conductivity of the zero temperature quantum critical theory ($T \to 0$).
The function (\ref{eq:acac}) that interpolates between these two constants describes the distribution of charged excitations in the system by energy scale, as we will review below.

In this paper we obtain the conductivities $\sigma_{ab}(\omega)$ in CFTs described by $U(N)$ Chern-Simons theories with fermionic or bosonic vector matter. The computations will be in the 't Hooft $N \to \infty$ limit, but will be exact in the finite coupling $\lambda = N/k$, with $k$ the Chern-Simons level (which is also taken to infinity). This exact treatment of a certain class of interactions is the main novel feature of our work. Previous analytic results on $\sigma_{ab}(\omega)$ have either employed a weakly interacting Boltzmann equation \cite{PhysRevB.56.8714, PhysRevB.57.7157} or else have used the strongly interacting framework of holography \cite{Herzog:2007ij}. One aspect of the problem we will not overcome is the resolution of the delta function in the $\sigma_{xx}$ conductivity at $\omega = 0$. As we shall explain, the delta function is due to the existence of (infinitely many) vectorial conserved charges in the $N = \infty$ theory that overlap with the total electric current operator $\vec J$. We will outline how the challenging incorporation of finite $N$ effects that resolve the delta function, while still working exactly in $\lambda$, might be possible using  memory matrix methods.

A remarkable fact about the Chern-Simons CFTs under study is that the theories with fermionic and bosonic matter are conjectured to be related by a strong-weak coupling duality \cite{Maldacena:2011jn, Maldacena:2012sf, Aharony:2012nh, Aharony:2012ns}. This is an instance of three dimensional bosonization. The ability to calculate exactly in the coupling $\lambda$ allows this duality to be explicitly probed.
Our results can be thought of as a nonzero temperature generalization of \cite{Aharony:2012nh, GurAri:2012is}, in which the zero temperature current-current correlators were obtained and the duality corroborated. We obtain the nonzero temperature conductivity of the fermionic theory at all couplings and verify that in a strong coupling expansion it is equal to the perturbative bosonic result. This new test of the duality (and check on our calculations) involves matching an entire function of $\omega/T$. In the opposite (weak coupling) limit, we also verified that our results reduce to those obtained for an Abelian Chern-Simons gauge field coupled to a fermion \cite{PhysRevB.57.7157}. Furthermore, we have resolved a discrepancy that was found in matching the $T=0$ current correlators in \cite{Aharony:2012nh, GurAri:2012is}.

Gapped non-Abelian Chern-Simons theories arise as the effective description of quantum Hall states with non-Abelian quasiparticles. For some older and newer results on this connection, see for instance \cite{Wen1, Frad1, Wen2, Seiberg:2016rsg, review, Dorey:2016mxm}. The most direct application of the non-Abelian Chern-Simons CFT results of this paper may be to quantum phase transitions out of these gapped non-Abelian states. At the critical point, charged degrees of freedom become gapless and couple to the Chern-Simons fields. In fact, quantum transitions out of general quantum Hall states, Abelian or non-Abelian \cite{chen, sondhi, maissam1,maissam2,nay}, involve strongly interacting gapless theories with broken parity. The ability to treat a class of interactions exactly makes the large $N$ Chern-Simons-matter CFTs we consider model theories for the study of general features of such interacting CFTs. Viewed as model condensed matter systems, the nonzero charge density dynamics of $U(N)$ Chern-Simons-matter theories has recently been explored \cite{Geracie:2015drf}. We will be considering the complementary effect of nonzero temperature at zero density.

\section{Theory and observables}

This section introduces the Chern-Simons-matter theories \cite{Giombi:2011kc, Aharony:2012nh,Aharony:2011jz}. The theories describe a $U(N)$ gauge field $A_\mu$, with Chern-Simons interactions, and a matter field in the fundamental representation of the gauge group (see Appendix~\ref{app:conventions} for our conventions).
They are defined on a two dimensional spatial plane and at temperature $T$.
The Chern-Simons action (in Euclidean time) at level $k$ is given by
\begin{align}
S_{\rm CS}(k) &= - \frac{ik}{8\pi} \int \! d^3x \, \epsilon^{\mu\nu\rho}
  \left( A^a_\mu \dho_\nu A^a_\rho +
  \frac{1}{3} f^{abc} A^a_\mu A^b_\nu A^c_\rho \right) \,. \label {SCS} 
\end{align}
Without matter, $k$ is an integer. We work in the 't Hooft limit $N,k \to \infty$, with the coupling 
\be
\lambda \equiv \frac{N}{k}
\ee
kept fixed. The fermionic theory includes Dirac fermions $\psi^i$, $i=1,\dots,N$, and the action is
\begin{align}
  S_{\rm fer} &= S_{\rm CS}(k) + 
  \int \! d^3x \, \bar{\psi} \gamma^\mu D_\mu \psi \label{Sfer}
\end{align}
where $D_\mu = \dho_\mu + A_\mu$.
The theory \eqref{Sfer} has a conserved $U(1)$ current given by 
\begin{align}
  J^\mu = i \bar{\psi} \gamma^\mu \psi \ed \label{Jfer}
\end{align}
Next, in the scalar theory we have complex scalars $\phi^i$, $i=1,\dots,N_b$ and the action is
\begin{align}
  S_{\rm sc} &= S_{\rm CS}(k_b) + 
  \int \! d^3x \left[ (D_\mu \phi)^\dagger D^\mu\phi
  + \frac{\lambda_4}{2 N_b} (\phi^\dagger\phi)^2 \right]
  \ed \label{Sscalar}
\end{align}
We take the limit $N_b,k_b \to \infty$ keeping the coupling $\lambda_b = N_b/k_b$ fixed.
To flow to the IR fixed point we take $\lambda_4 \to \infty$ (after taking the large $N_b$ limit) while tuning the zero-temperature scalar mass to zero. There is a conserved $U(1)$ current given by
\begin{align}
  J^{\bos}_\mu = i\phi^\dagger \Big( 
  \overleftarrow{D}_\mu - \overrightarrow{D}_\mu \Big) \phi
  \ed \label{Jbos}
\end{align}
Note that even though in (\ref{Jfer}) and (\ref{Jbos}), we have written down the local current, we will only be considering the total current throughout this paper, that is, the $k=0$ mode of the currents. All correlators below refer to the $k=0$ mode only.

The fermionic theory \eqref{Sfer} and the scalar theory \eqref{Sscalar} are conformal, and both have a marginal deformation that is parameterized by $\lambda$.
The fermionic and bosonic theories are conjectured to be dual to each other under the mapping \cite{Aharony:2012nh}
\begin{align}
  k_b = -k, \quad \lambda_b = \lambda - \sign(\lambda) \ed
  \label{map}
\end{align}
In our scheme, both 't Hooft couplings are understood to be in the range $|\lambda|,|\lambda_b| \le 1$.
Slightly beyond this range the theory is not unitary, because (for example) the stress-tensor 2-point function becomes negative \cite{Aharony:2012nh,GurAri:2012is}. The mapping (\ref{map}) takes a weakly coupled theory ($\lambda$ or $\lambda_b$ small) to a strongly interacting theory ($\lambda_b$ or $\lambda$ of order one).
Under the duality, the $U(1)$ currents \eqref{Jfer} and \eqref{Jbos} are mapped to each other.

The conductivities are two point functions of the currents \eqref{Jfer} or \eqref{Jbos}. The presence of the dynamical $U(N)$ Chern-Simons term in the theory enforces the operator equation
\begin{align}
  J^\mu = \frac{k}{4\pi} J^\mu_{\rm top} \ec
  \label{JJtop}
\end{align}
where $J^\mu_{\rm top} = \epsilon^{\mu\nu\rho} \, \text{tr} F_{\nu\rho}$ is a topological current associated to the diagonal $U(1)$ subgroup of $U(N)$.
Equation (\ref{JJtop}) describes the usual Chern-Simons dressing of each fermion with magnetic flux. That is to say, the constraint in (\ref{JJtop}) attaches magnetic flux to charged degrees of freedom \cite{zee1995quantum}. We could therefore equivalently be computing correlators of the topological current.

With the current operator \eqref{Jfer} or \eqref{Jbos} at hand, the conductivities $\sigma_{ab}$ can be obtained from the definition (\ref{eq:first}). The retarded Green's function of the currents is computed by analytic continuation of the Euclidean two point function from the upper half complex frequency plane to the real frequency axis, see e.g. \cite{forster}. Specifically, we will analytically continue the nonzero temperature Euclidean two point function $G_{J_a J_b}^R(i \omega_n) \equiv \langle J_a J_b\rangle(\omega_n)$, computed at general positive Matsubara frequency $\omega_n = -i\omega = 2\pi n T$ where $n \in \bZ^+$, a positive integer. Here $\{a,b\}=\{x,y\}$ are spatial indices.
We also denote $\rchi_{ab} = \langle J_a J_b\rangle(\omega_n=0)$, which is not in general equal to $G^R_{J_aJ_b} (\omega=0)$. 
From the analytically continued function $G_{J_a J_b}^R(z)$
the conductivity at real frequencies $\omega$ is, as announced in (\ref{eq:first}) above,
\be
\sigma_{ab}(\omega) = \lim_{z \to \omega + i 0^+} \frac{1}{i z} \left[G^R_{J_a J_b}(z) 
- \rchi_{ab}\right]
 \,.\label{sigma}
\ee
The $+ i 0^+$ reminds us that the analytic continuation is from the upper half complex frequency plane. This fact can be important, and corresponds to the physical requirement that the current be computed in the presence of a source that dies off in the far past.
In the derivation of Ohm's law via the Kubo formula, $\rchi_{ab}$ is also important; it appears as a surface term in a Laplace transform integral \cite{Luttinger:1964zz, forster}.
 
 It will be technically convenient to obtain the correlator $\langle J^+ J^-\rangle(\omega_n)$
 in `lightcone' coordinates $x^\pm = (x^1 \pm i x^2)/\sqrt{2} = (x\pm i y)/\sqrt{2}$.
In the theories we are considering there is a Chern-Simons term that breaks parity. Therefore the Hall conductivity $\sigma_{xy}$ is odd in $\lambda$ while the longitudinal conductivity $\sigma_{xx}$ is even.
From this consideration, it follows that the conductivities are related to the lightcone two point function as follows.
\begin{align}
  \sigma_{xx}(\omega) = 
  \sigma_{yy}(\omega) &= \lim_{z \to \omega + i 0^+} \frac{1}{i z} 
  \left[ G^R_{J^+ J^-}(z) - \rchi_{+-} \right]
 \Big\vert_{\lambda \mathrm{-even}}
  \ec \label{sigmaxx} \\
  \sigma_{xy}(\omega) = - \sigma_{yx}(\omega) &= \lim_{z \to \omega + i 0^+} \frac{1}{z} G^R_{J^+ J^-}(z)  \Big\vert_{\lambda \mathrm{-odd}}
  \ed \label{sigmaxy}
\end{align}
Here we used the fact that $\langle J^+ J^+ \rangle = \langle J^- J^- \rangle = 0$ due to rotational invariance. Throughout our paper, $\rchi$ will in fact be zero.
This can be verified by direct calculation, or deduced by gauge invariance \cite{zohar}: At high temperature one can dimensionally reduce the theory on the thermal circle. The resulting $2d$ theory is gapped. The Euclidean 2-point function at zero Matsubara frequency corresponds to a mass term in the effective action of the reduced gauge field, and must therefore vanish.

The conductivities defined above are direct probes of the charged excitations in the theory. Let us think clearly about what this means. The lightest gauge-invariant states that are charged under the $U(1)$ symmetry include one unit of magnetic flux and $k$ fundamental fermions, obeying equation \eqref{JJtop} \cite{Radicevic:2015yla, Aharony:2015mjs}. (The fundamental fermions themselves are not gauge-invariant and therefore do not correspond to charged states.)
In the 't Hooft limit we are considering here, in which $k \to \infty$, both the mass and charge of these states goes to infinity. These states therefore do not contribute to the conductivity at finite frequencies, as they are too heavy to be produced.
Instead the current will be carried by `electron-positron' pairs, moving in opposite directions (cf. \cite{PhysRevB.56.8714, PhysRevB.57.7157}) and joined by a Wilson line in order to be gauge-invariant.
We note in passing that one may consider instead a theory with light charged states by taking two fermion flavors $\psi^i_{\alpha}$, $i=1,\dots,N$, $\alpha=1,2$.
The theory then has an additional global $SU(2)$ flavor symmetry with light charged operators such as $\bar{\psi}_{1} \psi_2$.
It is easy to re-purpose the computations in this paper to this case.

\section{Physics and discussion of results}
\label{sec:physics}

Most of the computations in this paper will focus on the (slightly more tractable) fermionic theory. The two observables of interest, $\sigma_{xx}$ and $\sigma_{xy}$, are respectively even and odd in $\lambda$, from equations (\ref{sigmaxx}) and (\ref{sigmaxy}). This allows us to focus on the range of couplings $0 \leq \lambda \leq 1$. We will additionally perform a perturbative computation in the bosonic theory in order to match the fermionic result at $\lambda \approx 1$, and thereby corroborate the duality (\ref{map}).

\subsection{Dissipation and charged excitations}
\label{sec:diss}

The longitudinal and Hall conductivities are both complex functions. We will therefore be computing a total of four real functions of frequency. The most directly physical quantities are those that are necessarily positive in order for entropy production to be positive. According to the spectral representation of Green's functions, these quantities directly `count' the number of charged excitations in the theory as a function of energy, see e.g. \cite{forster,Hartnoll:2009sz}.

An external electric field creates a current according to Ohm's law: $j_a(\omega) = \sigma_{ab}(\omega) E^b(\omega)$. The average rate of work done on the system over a cycle (i.e. time period $2\pi/\omega$), per unit volume, is given by the Joule heating formula
\be
\dot w = \overline{E^a(\omega)} \chi''_{ab}(\omega) E^b(\omega) \,.
\ee
The dissipative matrix is given by
\be
\chi''(\omega) = \left(
\begin{array}{cc}
\text{Re} \, \sigma_{xx}(\omega) & i \, \text{Im} \, \sigma_{xy}(\omega)\\
- i \, \text{Im} \, \sigma_{xy}(\omega) & \text{Re} \, \sigma_{xx}(\omega)  
\end{array}
\right) \,. \label{eq:chipp}
\ee
This matrix is obtained from standard manipulations (e.g. \cite{forster,Hartnoll:2009sz}), together with the expression (\ref{eq:first}) for the conductivity.

The eigenvalues of the dissipative matrix must both be positive
\be\label{eq:spm}
\sigma_\pm(\omega) = \text{Re} \, \sigma_{xx}(\omega)  \pm \text{Im} \, \sigma_{xy}(\omega) \geq 0 \,.
\ee
These dissipative `eigenconductivities' are the response of the system to circularly polarized electric fields with the two possible chiralities. They are direct probes of the charged excitations in the system as a function of frequency. The first result we give is for these two functions of frequency. They are shown in the following Figure \ref{fig:sigmapm}.
\begin{figure}[h]
  \centering
  \includegraphics[width=0.7\textwidth]{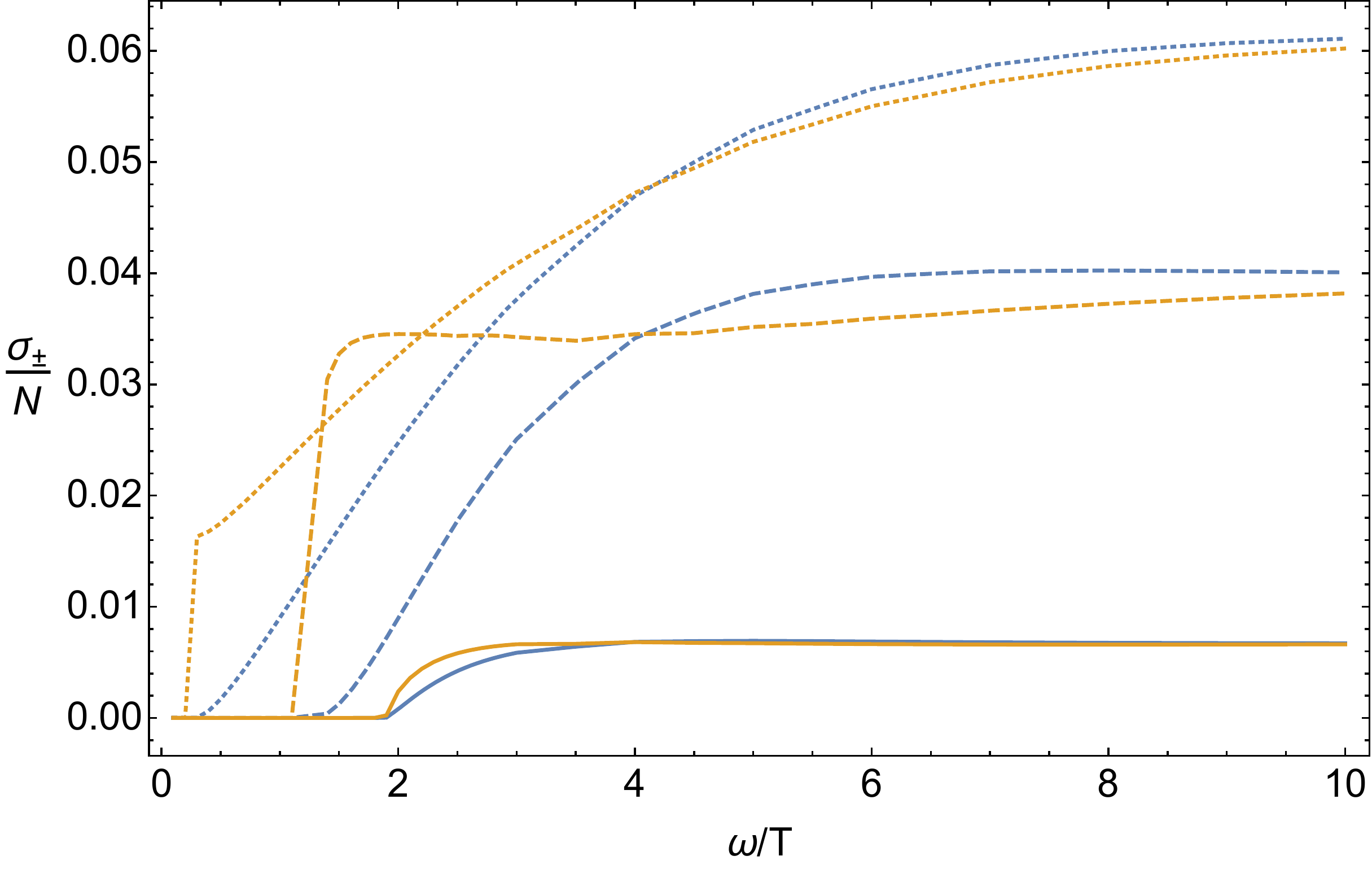}
  \caption{The dissipative conductivities $\sigma_\pm(\omega)$, defined in (\ref{eq:spm}), as a function of frequency in the fermionic Chern-Simons-matter theory. From top to bottom, the couplings are $\lambda = 0.1$ (dotted), $\lambda = 0.5$ (dashed) and $\lambda = 0.9$ (solid). In each pair of curves, the curve that rises fastest at low frequencies (orange) is $\sigma_+(\omega)$ and the other (blue) is $\sigma_-(\omega)$. These plots are missing a delta function at $\omega = 0$, as described in Section \ref{sec:div}.}
  \label{fig:sigmapm}
\end{figure}
The computations leading to these results will be described in later sections. The computations are analytic up to a final integral which is performed numerically.

The most distinctive feature of the plots in Figure \ref{fig:sigmapm} is that dissipation turns on only above some frequency, and is zero below. The threshold frequency is twice the thermal mass acquired by the fermions. The thermal mass is $m_F = \mu_F T$, with the dimensionless $\mu_F$ satisfying
\be
 \mu_F= \lambda \mu_F +
  \frac{2}{\pi} \, \text{Im} \left[
  \Li_2 \left( - e^{-\mu_F - i \pi \lambda} \right) \right] \label{muFresult} \,.
\ee
This equation for the thermal mass was previously derived in \cite{Aharony:2012ns}. The solution for $\mu_F$ is a monotonic function that goes from $\mu_F = 0$ (at $\lambda = 0$) to $\mu_F \approx 0.96$ (at $\lambda = 1$). The $\lambda=1$ value is simply the large $N$ thermal mass of the critical $O(N)$ model. Above the frequency $2 m_F$, `electron-hole' fermion pairs can be produced, leading to dissipation. The sharpness of the threshold is a consequence of the fact that to leading order in large $N$, the fermion self energy does not acquire an imaginary part (as is also seen in the large $N$ Wilson-Fisher fixed point \cite{PhysRevB.56.8714}). The fermion propagator will be discussed in more detail in subsequent sections.

At large frequencies, the curves in Figure \ref{fig:sigmapm} approach constants. This can happen because the conductivity is a dimensionless quantity in 2+1 dimensions. The limiting values are simply the $T=0$ values for the conductivities. These were previously obtained in \cite{Aharony:2012nh, GurAri:2012is}, up to a subtlety concerning the choice of regulator, that we will discuss later. The Hall conductivity is real at $T=0$ and hence,
\be
\left. \sigma_+ \right|_{T=0} = \left. \sigma_- \right|_{T=0} = \left. \sigma_{xx} \right|_{T=0} = \frac{N \sin(\pi \lambda)}{16 \pi \lambda} \,.
\ee
These values decrease monotonically as a function of coupling from a constant at $\lambda = 0$ to zero at $\lambda = 1$.

\subsection{Divergent dc conductivities and conserved operators}
\label{sec:div}

The plots in Figure \ref{fig:sigmapm} mask a key aspect of the $N = \infty$ physics. Namely, that there is a delta function in the dissipative conductivities $\sigma_\pm$ at $\omega = 0$. This delta function comes purely from $\text{Re} \, \sigma_{xx}(\omega)$. The weight $D$ of the delta function can be read off from the behavior of the imaginary part $\text{Im} \, \sigma_{xx}(\omega)$, which shows a characteristic $1/\omega$ dependence. This is because at low frequencies
\be\label{eq:delta}
\sigma_{xx}(\omega) = 
\frac{iD}{\pi} \lim_{\epsilon \to 0^+} \left( \frac{1}{\omega + i\epsilon} \right) + \cdots
= D \left(\delta(\omega) + \mathrm{p.v.} \frac{1}{\pi} \frac{i}{\omega} \right) + \cdots \,.
\ee
Plots of the real and imaginary parts of $\sigma_{ab}(\omega)$, in particular illustrating the $1/\omega$ behavior of $\text{Im} \, \sigma_{xx}(\omega)$ are given in Appendix \ref{app:allplots}.
We will refer to $D$ as the `Drude weight'. It is plotted in the following Figure \ref{fig:drude}. This plot is obtained by computing a certain integral numerically; the relevant integral will be given below.
\begin{figure}[h]
  \centering
  \includegraphics[width=0.7\textwidth]{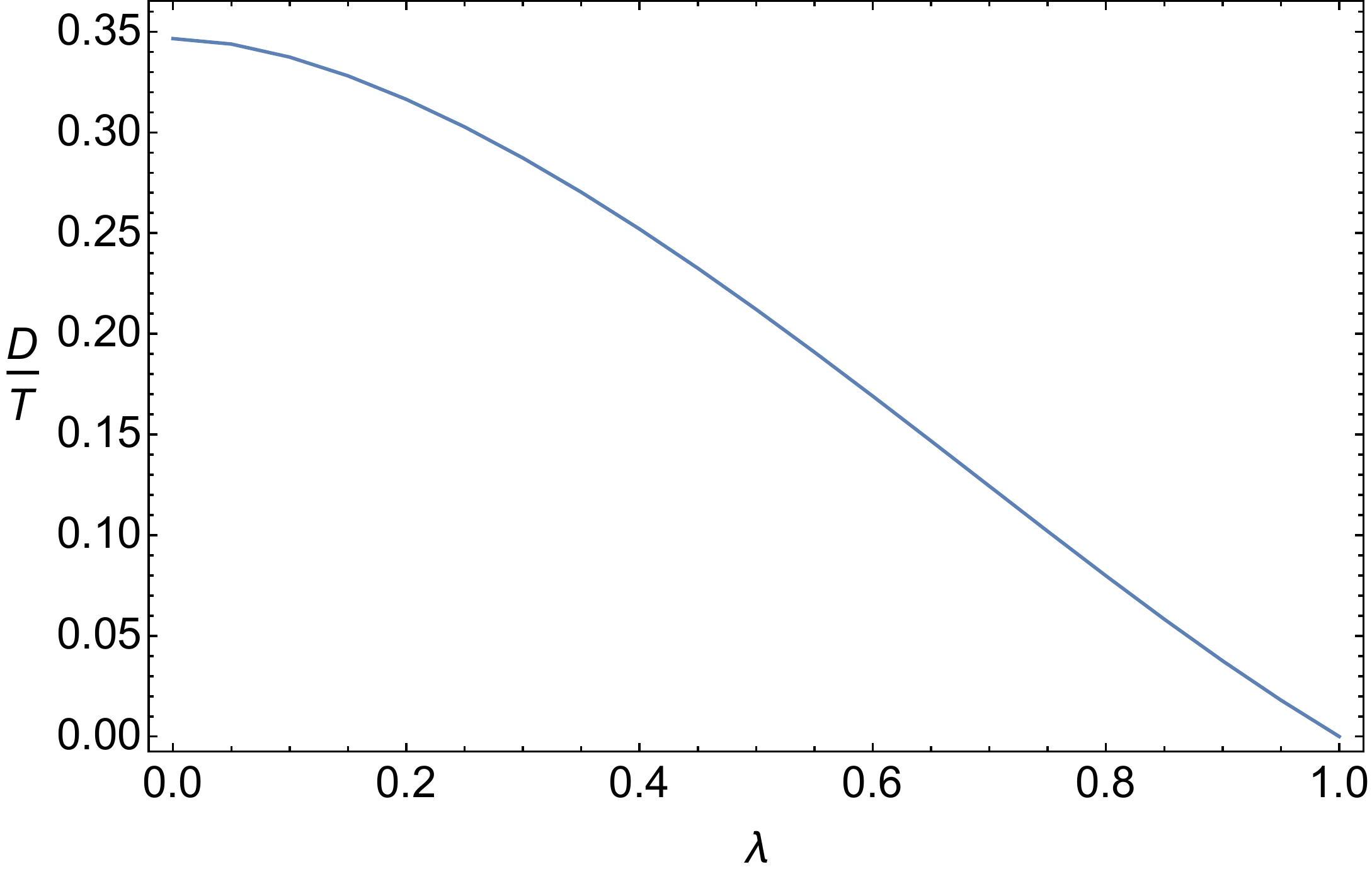}
  \caption{The Drude weight, $D$ in equation (\ref{eq:delta}), as a function of the coupling $\lambda$ in the fermionic theory. $D/T$ approaches $(\log 2)/2$ as $\lambda\to 0$, and $0$ as $\lambda \to 1$.}
  \label{fig:drude}
\end{figure}

From the definition of the conductivities (\ref{sigma}), the weight of the delta function satisfies 
\begin{align}
\frac{D}{\pi T} &= \frac{1}{T} \left[\rchi_{xx} - G^R_{J_xJ_x}(0) \right]\\
&= \frac{1}{V T^2 Z}\sum_{
  \begin{smallmatrix}
  m,n \\
  E_m=E_n
  \end{smallmatrix}
  }
 e^{-\beta E_n} 
\vert \langle n \vert J_{x} \vert m\rangle \vert^2 \ed
\label{eq-drudespectral}
\end{align}
Recall that $\chi_{ab} = \langle J_a J_b \rangle (\omega_n=0)$, whereas
$G^R_{J_aJ_b}(0)$ is defined via analytic continuation from the upper half complex frequency plane.
The second equality here can be derived by spectral representation of the various correlators.
A useful general discussion of such identities can be found in \cite{Ferrari:2016snh}.
(While $\rchi$ is zero in our case, it is not zero in general. For example, a free massless scalar will have 
$\rchi_{xx} \neq 0$.)
A delta function with the correct weight is crucial in order for our conductivities to satisfy the CFT sum rule \cite{Gulotta:2010cu, WitczakKrempa:2012gn, Katz:2014rla}
\be
\int_0^\infty \left(\text{Re} \, \sigma_{xx}(\omega) - \sigma_{xx}^\infty \right) = 0\,,
\ee
as we have checked. Here $\sigma_{xx}^\infty = \lim_{\omega \to \infty} \sigma_{xx}(\omega)$. The sum rule requires the leading operator appearing the current-current OPE to have dimension greater than one \cite{Katz:2014rla}. The leading such operator in our theory is the fermion mass operator, which has dimension two.

Damle and Sachdev \cite{PhysRevB.56.8714} emphasized that the delta function in the conductivity of nonzero-temperature CFTs is an artifact of working at $N = \infty$ in a vector model \cite{PhysRevB.56.8714}. If an electron and a positron move in opposite directions, the resulting state carries electric current but no momentum. If this electron-positron pair can decay, this will relax the current while preserving momentum. However, at $N=\infty$, as we have noted above and as we will see below, the fermionic propagators do not acquire an imaginary part in their self-energies. They are therefore stable and do not decay. These stable `quasiparticle' excitations can then continue to carry electric current once an external electric field is turned off, giving rise to an infinite conductivity. To resolve the delta function, $1/N$ effects such as the fermion lifetime must be computed and incorporated into the calculation of the conductivity.

We will now recast the Damle-Sachdev observation in a field-theoretic language. Namely, we will show how the delta function in (\ref{eq:delta}) is due to the overlap of the electric current operator with infinitely many operators originating from higher-spin currents that are almost conserved at $N = \infty$. This gives an inequality for the `Drude weight' as a sum over the operator overlaps. In Section \ref{sec:discuss} below, we suggest that this perspective may lead to a feasible framework for resolving the delta function by including certain $1/N$ effects while still working exactly in $\lambda$. If one is willing to work perturbatively in $\lambda$, the delta function can be resolved by solving a Boltzmann equation \cite{PhysRevB.56.8714, PhysRevB.57.7157}. However, the ability to work exactly in $\lambda$ is the exciting new feature of the Chern-Simons-matter theories under study.

In Appendix \ref{app:twopt} we adapt arguments by Mazur \cite{mazur} and Suzuki \cite{suzuki} to obtain the following
inequality, by expressing the Drude weight as the time-averaged correlation function of the total current (i.e. of the current operator at $\vec k = 0$),
\be
\frac{D T}{\pi} = \lim_{t_o \to \infty} \frac{1}{t_o} \int_{0}^{t_o}dt\, \langle J_x^\dagger(0) J_x(t) \rangle  \ge
  \sum_{a,b} 
  \langle J_x Q_a \rangle  C_{ab}
  \langle Q_b^\dagger J_x \rangle 
 \,. \label{eq:suz}
\ee
Here, angle brackets refer to the thermal expectation value at $\vec{k}=0$ (after stripping off a factor $\delta^{(2)}(0)$ that corresponds to the spatial volume), $Q_a$ are any set of constants of motion of the system and the matrix of overlaps $C_{ab}$ is defined as
\be
C_{ab} = \frac{ 
  \langle Q_a^\dagger Q_b \rangle
  }
  {
  \langle Q_a^\dagger Q_a \rangle
  \langle Q_b^\dagger Q_b \rangle
  } 
 \,.
\ee
The first equality between $D$ and the time-averaged correlator follows from 
(\ref{eq-drudespectral}) and a spectral decomposition for the time-averaged correlator,
see (\ref{eq-timeaveragespectral}).
More specifically, $Q_a$ must be constant inside thermal two point functions, so that for all single trace operators ${\mathcal O}$ we have
\be\label{eq:Tconserved}
\langle \dot Q_a(0) {\mathcal O}(t) \rangle  = 0 \,.
\ee
The result (\ref{eq:suz}) is essentially an instance of a Mazur inequality \cite{mazur, suzuki}. The right hand side is non-negative because the sum involves the positive-definite matrix $C_{ab}$.
 Therefore it is sufficient to find one conserved operator that overlaps with the current in order to demonstrate the presence of a delta function in the conductivity. The more such conserved operators that can be found, the stronger the bound on the Drude weight.

As we show in Appendix~\ref{app:twopt}, one such $Q_a$ operator can be constructed in the large $N$ Chern-Simons fermion theory from an almost-conserved high spin current.
We believe that infinitely many such $Q_a$ operators can be constructed in the large $N$ Chern-Simons fermion theory, all satisfying the conservation equation \eqref{eq:Tconserved}.
The starting point is the set of conserved high-spin currents $J^{(s)}_{\mu_1 \cdots \mu_s}$, with spins $s=1,2,3,\ldots$, of the free fermionic theory (\textit{i.e.} at $\lambda = 0$). While the full expression for the currents is complicated \cite{Giombi:2011kc}, the essential point is that
the spin $s$ current is given by $s-1$ covariant derivatives sandwiched between $\bar \psi$ and $\gamma \psi$. 
For each spin $s>1$ we can construct the following conserved constant of motion:
\be
Q^{(s)}_x = \int d^2x J_{(s)}^{tt\cdots tx}(x) \,.
\label{QmainText}
\ee
These operators will generally overlap with the electric current when $s$ is odd, leading to a non-trivial bound \eqref{eq:suz} in the free theory.

In the interacting theory ($\lambda \ne 0$) there are analogous currents $J^{(s)}$ that are not quite conserved, instead $\pa \cdot J^{(s)}$ is proportional to a multi-trace operator \cite{Giombi:2011kc}. In the absence of order $N$ vacuum expectation values for single trace operators, the effects of multi-trace operators inside two point functions are subleading at $N = \infty$. At $T > 0$, however, such expectation values are present. In Appendix \ref{app:twopt} we show how the spin 3 current operator $J^{(3)}$ can be improved by multi-trace operators at $\lambda > 0$, leading to a `sufficiently conserved' currents $\widetilde J^{(3)}$ even at $T>0$.
We then use the improved current to construct an improved constant of motion as in \eqref{QmainText}.
We believe that this construction can be generalized to other high-spin currents, leading to infinitely many improved constants of motion $Q^{(s)}_x$.

In this way, we finally obtain the desired lower bound on the Drude weight in terms of overlaps of the current operator with the conserved vector charges of the $N=\infty$ theory
\be\label{eq:dfinal}
\frac{D}{\pi T} \ge \frac{1}{T^2}
\sum_{
  \begin{smallmatrix}
  s,s'>1\\
 \text{odd}
  \end{smallmatrix}
  }
\langle J_x Q_x^{(s)} \rangle  C_{ss'}
  \langle Q_x^{(s')} J_x \rangle  \,.
\ee
We will not evaluate these overlaps explicitly. The point of our discussion here has been to give a clear field-theoretic picture for the delta function in the conductivity. The existence of these infinitely many conserved operators reflects the presence of underlying stable quasiparticles in the $N = \infty$ theory, even at $T>0$.

Before moving on, we emphasize that the physics of the delta function we have just discussed is different from that of the delta function in the conductivity of a system with a nonzero charge density $\langle J^t \rangle \neq 0$ and conserved momentum. The nonzero density delta functions are not artifacts of $N=\infty$ or weak interactions, but persist so
long as momentum is conserved \cite{Hartnoll:2007ih, Hartnoll:2012rj}. In the language developed above, this is because of the nonzero overlap $\langle J_x P_x \rangle \propto \langle J^t \rangle$. The physical picture is simple: With a net charge density, momentum necessarily carries current with it. But the momentum cannot relax if it is conserved, hence the current cannot relax (even in the absence of a driving electric field) and hence the conductivity is infinite. However, the CFTs we are considering here are at zero charge density, charge conjugation symmetry is preserved and hence the momentum cannot overlap with the electric current. Conserved momentum is not responsible for the delta function in (\ref{eq:delta}). In contrast, the infinite conductivity found in the same Chern-Simons-matter theory at nonzero charge density (and $T=0$) in \cite{Geracie:2015drf} is partially tied to momentum conservation. That divergence will not be fully broadened by the inclusion of a momentum-conserving quasiparticle decay at finite $N$ of the sort we describe below.

\subsection{Non-dissipative Hall conductivity}

So far we have discussed the real part of the longitudinal conductivity and the imaginary part of the Hall conductivity in Section \ref{sec:diss}, as these reveal the nature of dissipation in the nonzero temperature CFT. In Section \ref{sec:div} we used the low frequency behavior of the imaginary part of the longitudinal conductivity to extract the Drude weight. The real part of the Hall conductivity remains to be discussed.

\begin{figure}[ht]
  \centering
  \includegraphics[width=0.7\textwidth]{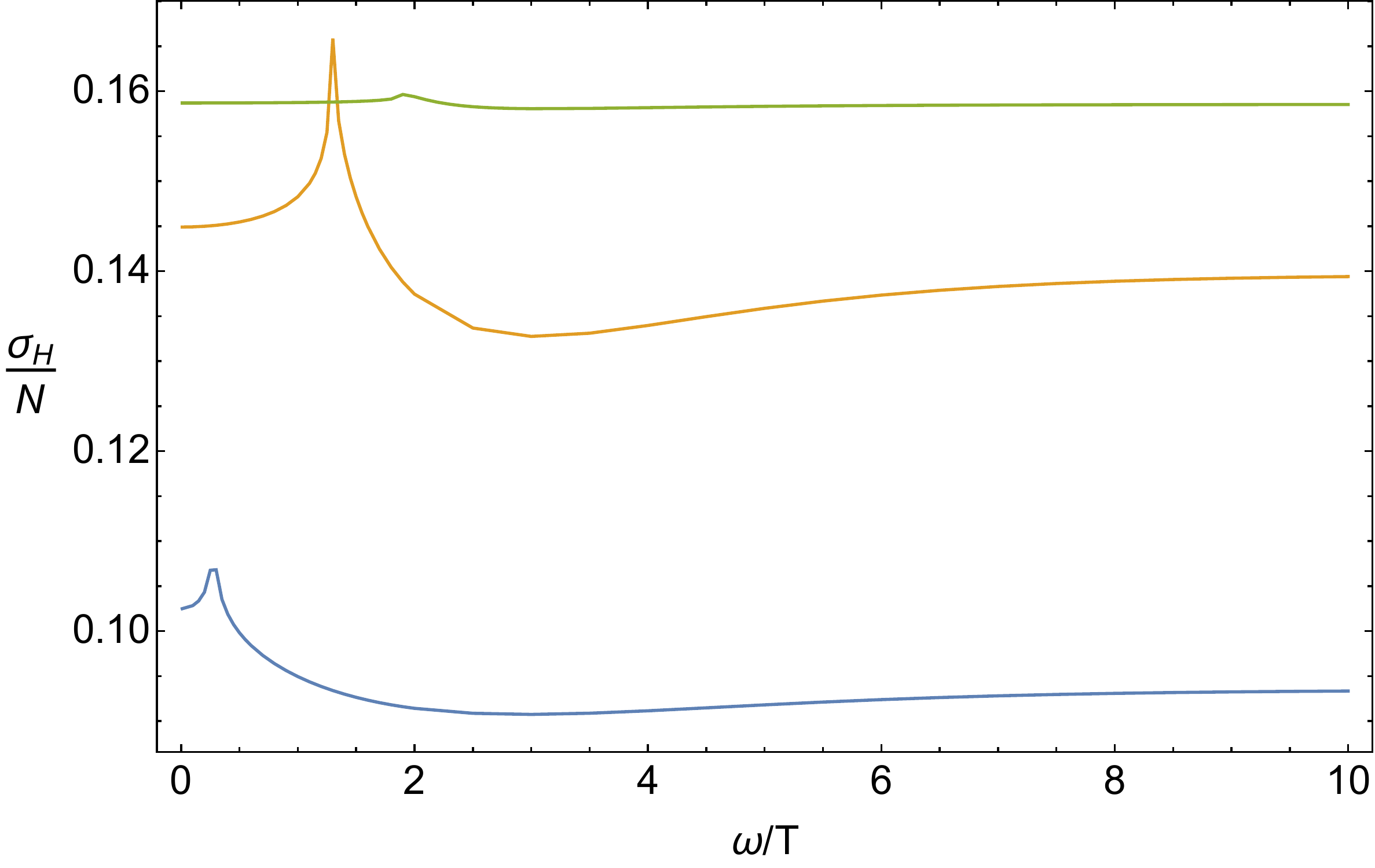}
  \caption{The non-dissipative Hall conductivities $\sigma_{H}(\omega)$, defined in (\ref{eq:sH}), as a function of frequency in fermionic Chern-Simons-matter theories. From bottom to top, the couplings are $\lambda = 0.1$, $\lambda = 0.5$ and $\lambda = 0.9$.}
  \label{fig:sigmaH}
\end{figure}

The real Hall conductivity,
\be\label{eq:sH}
\sigma_H(\omega) \equiv \text{Re} \, \sigma_{xy}(\omega) \,,
\ee
 is a non-dissipative observable. Unlike the other three conductivities just mentioned, it is finite and nonzero in both the limits $\omega \to 0$ and $T \to 0$. These limits characterize the motion of charge orthogonal to an applied constant electric field at finite and zero temperature, respectively. The whole function of $\omega/T$ gives basic universal data of the theory described by the CFT. Figure \ref{fig:sigmaH} shows plots of this function for three values of the coupling.
The plots show a cusp at twice the thermal mass and tend to finite, nonzero values at $\omega \to 0$ and $T \to 0$.  It is to be expected that some singular behavior is present at a frequency where particle-hole production onsets, as the non-dissipative conductivity is related to the dissipative imaginary part of the Hall conductivity through the Kramers-Kronig relation 
\be
\text{Re} \, \sigma_{xy}(\omega) = \frac{1}{\pi} \, \text{p.v.} \int \frac{\text{Im} \, \sigma_{xy}(\omega') d\omega'}{\omega' - \omega} \,.
\ee
The limiting values themselves are plotted in Figure \ref{fig:sigmaHdc} below. There is a simple monotonic dependence of $\sigma_H$ on the coupling in both limits. This is perhaps surprising given that the strongly coupled fermionic theory is a weakly interacting bosonic theory. Furthermore the behavior in the two limits, $\omega = 0$ and $T=0$ is quite similar.
\begin{figure}[H]
  \centering
  \includegraphics[width=0.7\textwidth]{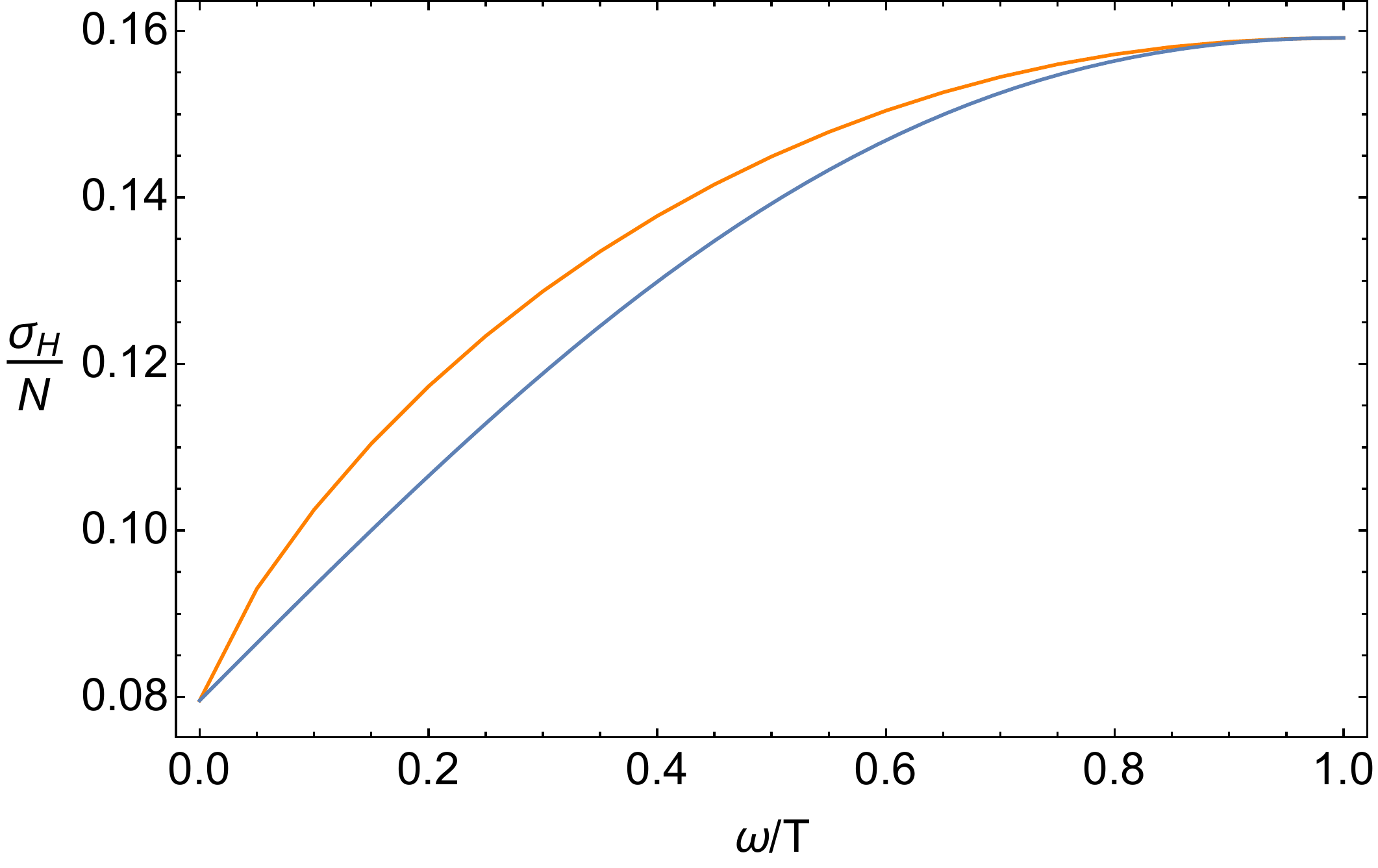}
  \caption{The non-dissipative Hall conductivities $\sigma_{H}$ at $\omega = 0$ (top) and $T = 0$ (bottom) as a function of the fermionic coupling $\lambda$.}
  \label{fig:sigmaHdc}
\end{figure}

Finally, in order to test the bosonization duality (\ref{map}) we can compare the strongly interacting fermionic conductivity, the top curve shown in Figure \ref{fig:sigmaH}, with a weakly interacting bosonic conductivity. The bosonic conductivity can then be simply obtained within perturbation theory. The results are shown in the following Figure \ref{fig:compare}.
\begin{figure}[h]
  \centering
  \includegraphics[width=0.7\textwidth]{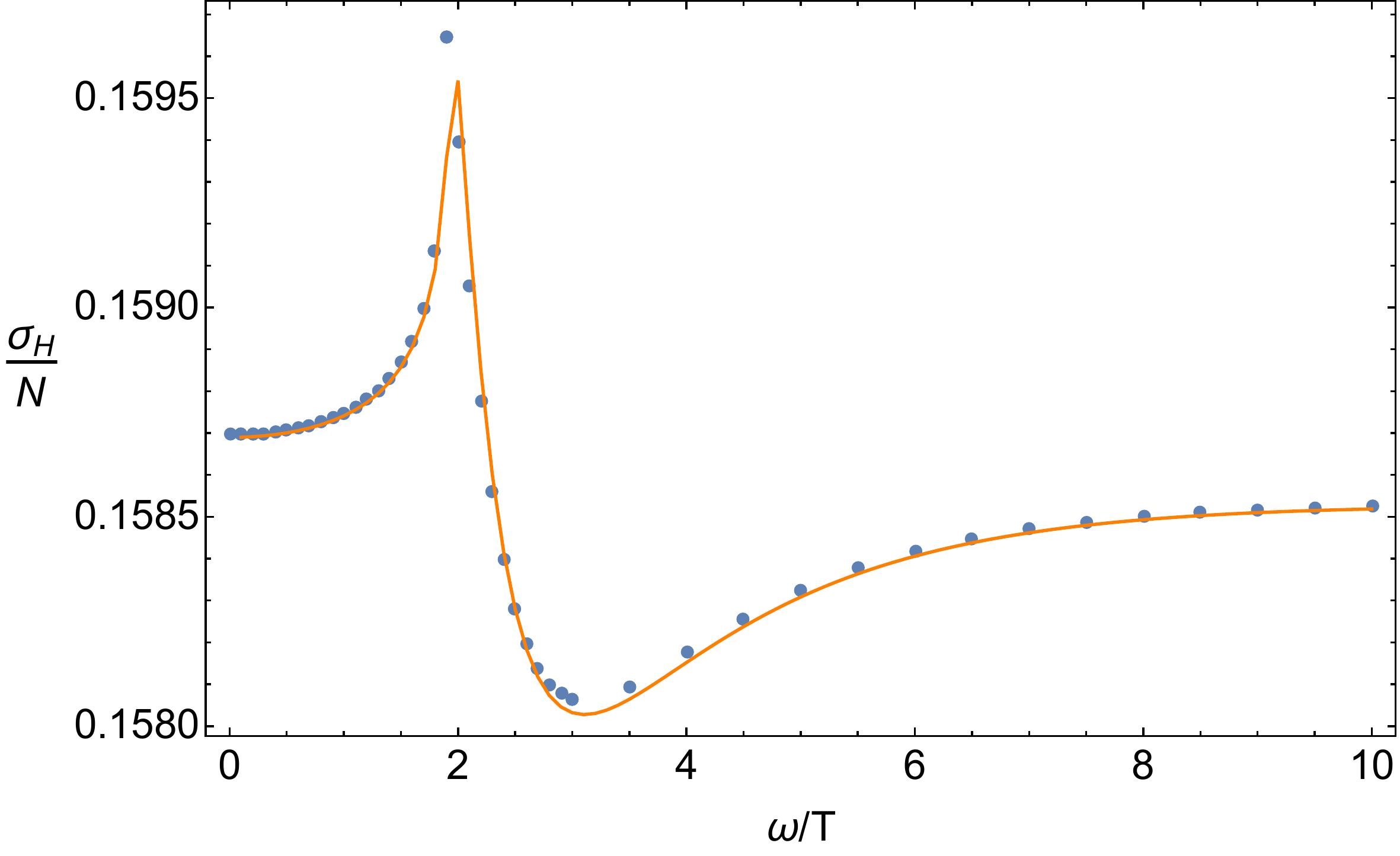}
  \caption{A functional test of the duality. The real part of the Hall conductivity of the fermionic theory at $\lambda=0.9$ (blue dots) compared with a perturbative calculation in the bosonic theory, evaluated at $\lambda_b = -0.1$ (orange curve).}
  \label{fig:compare}
\end{figure}
The agreement in Figure \ref{fig:compare} is seen to be excellent, giving a whole function's worth of corroboration to the duality. We will furthermore see that an expansion of the fermionic result about $\lambda = 1$ precisely reproduces the perturbative bosonic result.

\subsection{Integral expressions for the conductivities}
\label{sec:integrals}

The plots presented above have been obtained by numerically evaluating certain integrals. The integrals are obtained by solving Schwinger-Dyson equations for elements of the current-current correlator. Before giving the derivation of the results, we will write down these integral expressions for the conductivity. These are the main analytic results of our work. The full large $N$, exact in $\lambda$ current-current two-point function in the fermionic theory is given below in \eqref{JJfer}. 
The longitudinal and Hall conductivities are extracted from this two-point function via 
\eqref{sigmaxx} and \eqref{sigmaxy}.  According to those formulae, the basic ingredients are the parts of the correlator that are even and odd in $\lambda$.

The $\lambda$-even part of the 2-point function is
\begin{align}
  \langle J^+ J^-(\omega_n) \rangle_{\lambda\mathrm{-even}} &=
  - \frac{N}{8\pi^2 \lambda \beta} \lim_{\Lambda' \to \infty}
  \int_0^{\Lambda'} dx \, \Bigg\{ 2\pi \lambda + x F(n,x) \times \nonumber \\
&  \Bigg[
  \frac{(x^2 + \mu_F^2 - \pi^2 n^2)x f(x)}{\pi n}  
  \sinh \left( n \int_x^{\Lambda'} \! dy \, y F(n,y) \right)
  \cr & \qquad \qquad \;\;
  - i (x^2 + \mu_F^2 + x^2 f^2(x))
  \cosh \left( n \int_x^{\Lambda'} \! dy \, y F(n,y) \right)
 \Bigg]
  \Bigg\} \ed
\end{align}
Here and below $\Lambda' = \beta \Lambda$ where $\Lambda$ is the UV cutoff on momentum in the $x,y$ directions.  The sums over Matsubara frequencies have already been performed and are finite, requiring no regularization.

The $\lambda$-odd 
part of the current-current two-point function is given by
\begin{align}
  \langle J^+ &J^-(\omega_n) \rangle_{\lambda\text{-odd}} = 
  \frac{N}{2}i\omega_n + \frac{N\lambda}{8\pi} i\omega_n
  -\frac{N}{8\pi^2 \lambda \beta} \lim_{\Lambda' \to \infty}
  \int_0^{\Lambda'} dx \, x F(n,x) \times
  \cr
  &\Bigg\{
  2\pi n x f(x) + 
  \frac{x f(x) (x^2 + \mu_F^2 - \pi^2 n^2)}{\pi n}
  \left[ 1 - \cosh \left(
  n \int_x^{\Lambda'} \! dy \, y F(n,y)
  \right) \right]
  \cr & \quad
  + i (x^2 + \mu_F^2 + x^2 f^2(x)) \sinh \left( 
  n \int_x^{\Lambda'} \! dy \, y F(n,y)
  \right)
  \Bigg\} \ed
\end{align}
The two functions used in the above expressions are defined as follows:
\begin{align}
  x f(x) &= -\lambda \sqrt{x^2 + \mu_F^2}
  + \frac{2}{\pi} \text{Im} \, \Li_2 \left( - e^{-\sqrt{x^2 + \mu_F^2} + \pi i \lambda} \right)  \ec \label{fresult}\\
F(n,x) &=
  \frac{
  2 \log\cosh [ (\sqrt{x^2 + \mu_F^2} - \pi i \lambda)/2 ]
  - \cc}
  {\sqrt{x^2 + \mu_F^2} (\pi^2 n^2 + x^2 + \mu_F^2)} \label{Fresult} \ed
\end{align}
Recall that $\mu_F$ was defined in (\ref{muFresult}).

For the purposes of checking the duality, we will furthermore compute the current-current correlator in the bosonic theory perturtabtively at small $\lambda_b$. We find
\begin{align}
  \langle J_b^+ J_b^-(\omega_n) \rangle &=
  -\frac{N_b \lambda_b}{8\pi} i\omega_n - \frac{N_b (\pi^2 n^2 + \mu_0^2)}{4\pi\beta}
  \int_{\mu_0}^{\infty} dx
  \frac{\coth(x/2)}{x^2 + \pi^2 n^2} \nonumber \\
  &
  - \frac{iN_b\lambda_b n (\pi^2 n^2 + \mu_0^2 )}{4\beta}
  \left[ \int_{\mu_0}^{\infty} dx 
  \frac{\coth(x/2)}{x^2 + \pi^2 n^2}
  \right]^2 + O(\lambda_b^2) \,.
  \label{JJscintro}
\end{align}

The fermionic and bosonic theories are conjectured to be dual to each other under the mapping \eqref{map}.
Expanding the fermionic result near $\lambda=1$, leads to a precise agreement between the fermionic and bosonic theories. The pieces of the correlator coming from the anomaly -- as we will describe shortly -- are crucial for the correlators to match. This constitutes a new test of the bosonization duality at finite temperature.

\section{Zero temperature Hall conductivity}
\label{sec:zeroT}

Before working at nonzero temperatures, we need to understand a subtlety concerning anomalies and regularization that is present already at $T=0$, but which was not addressed in earlier works \cite{GurAri:2012is,Aharony:2012nh}. Correctly resolving this point leads to an extra constant (frequency-independent) contribution to the Hall conductivity.

In a CFT at nonzero temperature $T$ we have noted that the conductivities are a function of $\omega/T$, where $\omega$ is the source frequency. The zero temperature limit is therefore the same as the high frequency limit.
At $T=0$, the current 2-point function takes the general form 
\begin{align}
  \langle J^\mu J^\nu(p) \rangle =
  \tau \cdot
  \frac{p^\mu p^\nu - \delta^{\mu\nu} p^2}{|p|}
  + \frac{\kappa}{2\pi} \cdot
  \epsilon^{\mu\nu\rho} p_\rho \ed
  \label{CFTJJ}
\end{align}
These 2-point functions were computed in \cite{GurAri:2012is,Aharony:2012nh} for the fermionic and scalar theories.
The longitudinal conductivity is the same in both cases, and is given by
\begin{align}
  \left. \sigma_{xx} \right|_{T=0} = \tau 
  = \frac{N\sin(\pi\lambda)}{16\pi\lambda}
  \ed
\end{align}
Here $N,\lambda$ refer to either the fermionic or bosonic $N,\lambda$, according to the theory in which one is working. That is, this formula is invariant under the duality (\ref{map}). As with many of our formulae, upon putting $N=1$ it agrees to first order in perturbation theory with the result from Abelian Chern-Simons theory \cite{PhysRevB.57.7157}.
The Hall conductivity is given by the second term in \eqref{CFTJJ}: 
$\left. \sigma_{yx} \right|_{T=0} = \kappa/2\pi$.
The contribution to $\kappa$ from summing planar diagrams in our scheme was computed in \cite{GurAri:2012is,Aharony:2012nh}.
For the fermion theory it is given by
\begin{align}
  \frac{k}{4} \sin^2 \left( \frac{\pi\lambda}{2} \right) \,.
  \label{kferdyn}
\end{align}
For the scalar theory the result is similarly $\frac{k_b}{4} \sin^2 
\left( \frac{\pi\lambda_b}{2} \right)$.
As we shall now explain, there are additional contributions that must be included in order to cancel an anomaly in the background $U(1)$ symmetry.
These contributions were not fully considered in previous works on large $N$ Chern-Simons-matter theories, but they are important for obtaining the correct conductivity.

Let us begin by reviewing the properties of the Hall conductivity term at zero temperature \cite{Closset:2012vp}, namely the second term in \eqref{CFTJJ}.
This is a contact term, proportional to $\epsilon^{\mu\nu\rho} \dho_\rho \delta(x)$ in spacetime coordinates.
Such terms can often be shifted by introducing terms to the action that involve only the background fields.
In our case, let $a_\mu$ be the background Abelian gauge field that couples to the current ($a_\mu$ is real).
Then we can shift $\kappa \to \kappa + \delta\kappa$ by introducing the following term:
\begin{align}
  \frac{i\delta\kappa}{4\pi}
  \epsilon^{\mu\nu\rho}
  \int \! d^3x \, a_\mu \dho_\nu a_\rho \ed
  \label{bgCS}
\end{align}
We stress that this is a background term that is independent of the dynamical Chern-Simons term \eqref{SCS}.
Suppose we start with a theory that is gauge invariant in the background $U(1)$ symmetry.
In order to preserve this invariance we must choose the level $\delta\kappa$ to be an integer.\footnote{
This is the correct quantization when the minimal $U(1)$ charge in the system is 1, as it is in our case.
}
Therefore, the fractional part of $\kappa$ is a universal physical observable (one that is independent of our scheme), while its integer part is scheme-dependent \cite{Closset:2012vp}.
On the other hand, if the background gauge symmetry is anomalous then we may introduce a background term that is not properly quantized to cancel the anomaly.
A well-known example is the parity anomaly \cite{Redlich:1983dv}: The theory of a free Dirac fermion in $3d$ is anomalous, and the anomaly can be canceled by introducing a background Chern-Simons term \eqref{bgCS} with a half-integer level $\delta\kappa$.

One way to check whether the symmetry is anomalous is to give all matter fields a mass and flow to the IR by taking the mass to infinity.
In this limit one obtains an effective action for the background field.
If the Chern-Simons level of this effective action is not properly quantized, the symmetry is anomalous.

Let us consider the fermion theory with a mass term $\sigma \bar{\psi} \psi$ for the fermion.
We compute the current 2-point function in this theory in Appendix~\ref{app:zeroTmassanomaly}.
Taking the IR limit $\sigma/|p| \to \pm \infty$, we find that
\begin{align}
  \langle J^\mu J^\nu(p) \rangle \to
  \frac{1}{2\pi} \left( 
  \sign(\sigma) \frac{N}{2} - \frac{N\lambda}{4}
  \right)
  \epsilon^{\mu\nu\rho} p_\rho \ed
  \label{anomaly}
\end{align}
The first term is the usual parity anomaly contribution, but the second term is new.
Such a term cannot appear in a gauge-invariant scheme, but our regulator is a hard cutoff which breaks gauge invariance.
It is therefore necessary in this case to restore gauge invariance by adding an appropriate counter term.
In order to cancel the anomaly \eqref{anomaly}, we introduce a counter term of the form \eqref{bgCS} with
$\delta \kappa = \frac{N}{2} + \frac{N\lambda}{4}$.
This term shifts the value of the contact term $\kappa$ in the massless fermion theory.
The 2-point function in this theory is then given by \eqref{CFTJJ} where
\begin{align}
  \kappa_{\rm fermion} =
  \frac{k}{4} \sin^2 \left( \frac{\pi\lambda}{2} \right)
  + \frac{N}{2} + \frac{N\lambda}{4} \ed
  \label{kappafer}
\end{align}
Here we included the planar diagram calculation \eqref{kferdyn} and the anomaly cancelling piece $\delta\kappa$.

A similar calculation in the scalar theory with $U(N_b)$ gauge group at level $k_b$ leads to the result
\begin{align}
  \kappa_{\rm scalar} = 
  \frac{k_b}{4} \sin^2 \left( \frac{\pi\lambda_b}{2} \right)
  - \frac{N_b\lambda_b}{4} \ed
  \label{kappasc}
\end{align}
The $O(\lambda)$ parts of \eqref{kappafer} and \eqref{kappasc} agree with previous perturbative calculations of the Hall conductivity in Abelian Chern-Simons theories coupled to fermions and scalars \cite{PhysRevB.57.7157,spiridonov1991two}.

The fractional parts of $\kappa_{\rm fermion}$ and $\kappa_{\rm scalar}$ are scheme-independent observables.
They should therefore agree under the bosonization duality \eqref{map}, 
and it is easy to check that this is indeed the case.
This constitutes a new test of the bosonization duality at zero temperature.
Note that the anomaly-canceling counter-terms were crucial in obtaining this agreement. In our computations below, we must also add these counter-terms to the finite temperature Hall conductivities.

The limit $\lambda \to 1$ at large but fixed $N$ gives another check on the results above.
In this limit the stress tensor 2-point function vanishes, and the theory becomes topological (or empty).
Therefore in this limit the anomaly should vanish, and indeed we see that $\kappa_{\rm fermion}$ becomes an integer.
Similarly, the scalar anomaly vanishes in the limit $\lambda_b \to 1$ at fixed $N_b$.

\section{Nonzero temperature conductivity}

Let us now turn to the computation of the conductivities at nonzero temperature, the results of which were discussed in Section \ref{sec:physics}.
In the fermionic theory we write down the
exact answer in the 't Hooft coupling $\lambda$, while in the bosonic theory we compute to leading order in the coupling $\lambda_b$. 
We find that the scalar result agrees with the fermion conductivity in the strong $\lambda$ coupling limit, providing a new test of the bosonization duality.

\subsection{The role of non-trivial holonomy}
\label{sec:holonomy}

Gauge theories at nonzero temperature can have non-zero holonomy along the thermal cycle, which we can write schematically as $\langle \int_0^\beta dx^3 A_3 \rangle \ne 0$. In particular, this is known to be the case for our Chern-Simons matter theories \cite{Aharony:2012ns}. This holonomy affects physical observables such as the thermal free energy.
In an infinite volume limit ($T^2 V_2 \gg N$, where $V_2$ is the spatial volume), the holonomy can be computed explicitly and taken into account in calculations; we will work in this limit.
Let us now briefly review how the holonomy is included in finite temperature calculations \cite{Aharony:2012ns}.

The field $A_3(x)$ consists of the holonomy, which can be treated as a constant background $A_3^{\rm back}$ in the large $N$ limit, and a fluctuating part over which we integrate as usual.
The background holonomy is a diagonal matrix in color space that is independent of $x$. The diagonal elements can be written as
\begin{align}
  (A_3)_{ii}^{\rm back} = \frac{2\pi i}{\beta} \alpha_{ii} \ecq i=1,\dots,N \,.
\end{align}
In the large $N$ limit the discrete elements go over to a smooth distribution which is given by
\begin{align}
  \alpha_{ii} \to \alpha(u) = -|\lambda|u \ecq
  u \in [-1/2,1/2] \ec
\end{align}
and traces over color turn into integrals,
\begin{align}
  \frac{1}{N} \sum_{i=1}^N f(\alpha_{ii}) \to \int_{-1/2}^{1/2} du f(\alpha(u)) \ed
  \label{holonomy-integral}
\end{align}
Loop momenta are shifted in the usual way by $k_\mu \to \tilde{k}_\mu \equiv k_\mu - i A_\mu^{\rm back}$ so we have
\begin{align}\label{eq:shift}
  \tilde{k}_{\pm} = k_{\pm} \ecq
  \tilde{k}_3 = k_3 + \frac{2\pi}{\beta} \alpha(u) \ed
\end{align}
 
\subsection{Exact computation in the fermionic theory}

Computing the current-current correlator exactly in $\lambda$ in the theory \eqref{Sfer} at large $N$ involves taking an infinite sum of planar diagrams. 
The sum can be expressed as a combination of fermion propagator corrections and vertex corrections (the gluon propagator is not corrected in the large $N$ limit).
Each of these can be handled separately by solving an appropriate Schwinger-Dyson (SD) equation.

\begin{figure}[h]
  \centering
   \includegraphics[width=0.3\textwidth]{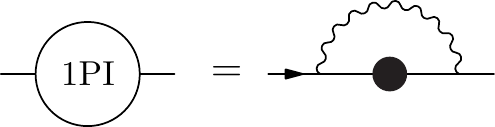}
  \includegraphics[width=0.8\textwidth]{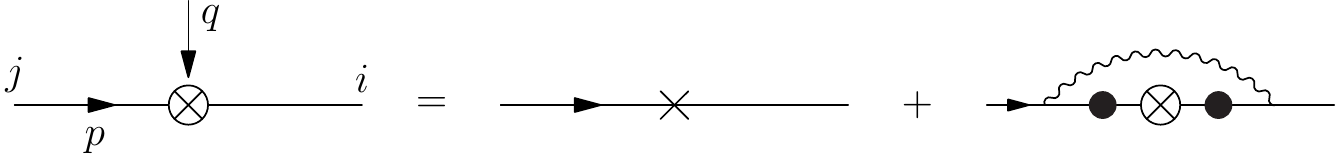}
  \caption{Schwinger-Dyson equations for the fermion propagator and the current vertex.
  Solid lines denote tree-level fermion propagators, and wiggly lines denote gluon propagators.
  An $\times$ denotes the tree-level current insertion (with a polarization $\mu$ that is not shown), an $\otimes$ denotes the exact vertex insertion, and a black circle denotes the exact fermion propagator.
  }
  \label{fig:ferSD}
\end{figure}

First, one computes the exact fermion propagator by solving the Schwinger-Dyson equation shown in Figure \ref{fig:ferSD}. The exact finite temperature propagator was already computed in \cite{Aharony:2012ns}.
It is given by
\begin{gather}
  \langle \psi_i(p) \bar{\psi}_j(-k) \rangle =
  \delta^j_i S_{(i)}(p) \cdot (2\pi)^3 \delta(p-k) \ec \cr
  S_{(i)}(p) \equiv 
  \frac{-i \tilde{p}_\mu^{ii} \gamma^\mu
  + i g(\beta p_s) p_+ \gamma^+ - f(\beta p_s) p_s I}
  {(\tilde{p}_{ii})^2 + \beta^{-2} \mu_F^2} \ed \label{fermionpropagatorexact}
\end{gather}
Here $\mu_F$ is the thermal mass of the fermion, whose sign is equal to $\sign(\lambda)$.
The thermal mass $\mu_F$ and the functions $f(x)$ and $g(x)$ all depend on $\lambda$, and are defined as follows.
\begin{align}
  \mu_F &= \lambda \mu_F +
  \frac{1}{\pi i} \left[ 
  \Li_2 \left( - e^{-\mu_F - \pi i \lambda} \right) - \cc
  \right] \label{muF} \ec \\
  x f(x) &= -\lambda \sqrt{x^2 + \mu_F^2}
  + \frac{1}{\pi i} \left[ 
  \Li_2 \left( - e^{-\sqrt{x^2 + \mu_F^2} + \pi i \lambda} \right) - \cc
  \right] \ec \label{f} \\
  x^2 g(x) &= x^2 f^2(x) - \mu_F^2 \label{gf} \ed
\end{align}

We proceed by computing the exact vertex
\begin{align}
  \langle J^\mu(-q) \psi_i(k) \bar{\psi}^j(-p) \rangle =
  V^\mu(q,p) \delta^j_i \cdot (2\pi)^3 \delta(q+p-k) \ed
  \label{Vdef}
\end{align}
We set the spatial momentum to zero ($q^\pm = 0$) and choose $q^3 = \frac{2\pi m}{\beta}$ where $m \in \bZ$ and $m>0$.
From now on we will simply use $q$ to refer to $q^3$.
The vertex at zero temperature was computed in \cite{GurAri:2012is} and we now extend this computation to nonzero temperature.

\subsubsection{Computation of the $T>0$ current vertex}

The Schwinger-Dyson equation for the vertex is shown in Figure \ref{fig:ferSD} and can be written as
\begin{align}
  V^\mu(q,p) = i\gamma^\mu - \frac{2\pi i \lambda}{\beta}
  \int_{-1/2}^{1/2} du
  \int \frac{d^2k}{(2\pi)^2} 
  \frac{1}{(k-p)^+}
  \sum_{n=-\infty}^{\infty}
  \gamma^{[3|}
  S(k+q) V^\mu(q,k) S(k) \gamma^{|+]} \ed
  \label{bs}
\end{align}
The anti-symmetrization of the $\gamma$'s is unweighted.

We will compute only the $V^+$ vertex as this is enough 
to compute $\langle J^+ J^- \rangle$.
To simplify the right hand side of \eqref{bs} we use the fact that for a $2 \times 2$ matrix $A = a_\mu \gamma^\mu + a_I I$ we have $\gamma^{[3|} A \gamma^{|+]} = 2 a_I \gamma^+ - 2 a_- I$.
Noting that only $\gamma^+$ and $I$ appeared in this last expression, we can write
\begin{align}
  V^+(q,p) &= V^+_+(q,p) \gamma^+ + V^+_I(q,p) I \ed 
\label{V++VI+}
\end{align}

Working out the spinor algebra in \eqref{bs}, we find the following equations:
\begin{align}
  V^+_+(q,p) &= i - \frac{4\pi i \lambda}{\beta}
  \int_{-1/2}^{1/2} du \int \frac{d^2k}{(2\pi)^2} \sum_{n=-\infty}^\infty
  \frac{1}
  { (k-p)^+ [\tilde{k}^2 + T^2 \mu_F^2] [(\tilde{k}+q)^2 + T^2 \mu_F^2] }
  \cr &\quad \times
  \left\{
  \left[ -\tilde{k}_3^2 + (f^2(k)+g(k)-1) k_s^2 - \tilde{k}_3 q_3 \right]
  V_I(q,k) + k^+ (2if(k) k_s - q_3) V^+_+(q,k)
  \right\} \ec \cr
  \label{V++}\\
  V^+_I(q,p) &= - \frac{4\pi i \lambda}{\beta}
  \int_{-1/2}^{1/2} du \int \frac{d^2k}{(2\pi)^2} \sum_{n=-\infty}^\infty
  \frac{1}
  { (k-p)^+ [\tilde{k}^2 + T^2 \mu_F^2] [(\tilde{k}+q)^2 + T^2 \mu_F^2] }
  \cr &\quad \times
  k^+ \left[ 2k^+ V^+_+(q,k) - (2if(k) k_s + q_3) V^+_I(q,k) \right] \ed
  \label{VI+}
\end{align}
Note that $p_3$ doesn't appear on the right hand side of these equations, so $V^+$ is independent of $p_3$.
Also, due to rotation invariance in the lightcone plane and from dimensional analysis, we can write
\begin{align}
  V^+_+(q,p) = v_+(m,y) \ecq
  V^+_I(q,p) = \beta p^+ v_I(m,y) \ed
  \label{Vv}
\end{align}
Here and below we will use the notation $x = \beta k_s$ and $y = \beta p_s$ (see Appendix~\ref{app:conventions} for our conventions).
We first carry out the sums over the thermal modes, which are given by
\begin{align}
  &\sum_{n=-\infty}^\infty
  \frac{1}{ [\tilde{k}^2 + T^2 \mu_F^2] [(\tilde{k}+q)^2 + T^2 \mu_F^2] }
  = 
  \frac{\beta^4
  \left[ 
  \tanh \left( \frac{1}{2} \sqrt{x^2 + \mu_F^2} - \pi i \alpha(u) \right)
  + \cc
  \right]
  }{8 \sqrt{x^2 + \mu_F^2}
  \left( \pi^2 m^2 + x^2 + \mu_F^2 \right)}
  \ec
  \label{sum1}
\end{align}
\begin{align}
  &\sum_{n=-\infty}^\infty
  \frac{\tilde{k}_3 (\tilde{k}_3 + q_3)}
  { [\tilde{k}^2 + T^2 \mu_F^2] [(\tilde{k}+q)^2 + T^2 \mu_F^2] }
  = 
  \frac{\beta^2 \sqrt{x^2 + \mu_F^2}
  \left[ 
  \tanh \left( \frac{1}{2} \sqrt{x^2 + \mu_F^2} - \pi i \alpha(u) \right)
  + \cc
  \right]
  }{8 (\pi^2 m^2 + x^2 + \mu_F^2)} \ed
  \label{sum2}
\end{align}
It will be convenient to define the combination 
$\tilde{x} = \sqrt{x^2 + \mu_F^2}$.
Let us also define
\begin{align}
 \tilde{x} (\pi^2  m^2 + \tilde{x}^2) F(m,x) &= - 2\pi i \lambda
  \int_{-1/2}^{1/2} du\,\,
  \text{Re}\,\tanh\, (\tilde{x}/2 + \pi i \vert \lambda \vert u) \cr &=
  2 \log\cosh\, [(\tilde{x} - \pi i \lambda)/2 ]
   - \cc \,.
  \label{F}
\end{align}
Next we compute the holonomy integral in the SD equations \eqref{V++}, \eqref{VI+}.
The result is
\begin{align}
  p^+ v_I(m,y) &= \beta^2 \int \frac{d^2k}{(2\pi)^2}
  \frac{(k^+)^2}{(k-p)^+} \left[ 
  v_+(m,x) - (\tilde{f}(x) + \pi m) v_I(m,x)
  \right] F(m,x) \ec
  \label{inteq1}\\
  v_+(m,y) &= i + \beta^2 \int \frac{d^2k}{(2\pi)^2}
  \frac{k^+}{(k-p)^+} \left[ 
  (g(x)-1) x^2 v_I(m,x)
  + ( \tilde{f} - \pi m) v_+(m,x)
  \right] F(m,x) \ec
  \label{inteq2}
\end{align}
where we have defined another shorthand,
\begin{align}
  \tilde{f}(x) = i x f(x) \ed
\end{align}
One can now carry out the angular integral in the spatial plane (for example using contour techniques) and get
\begin{align}
  v_I(m,y) &= \frac{1}{2\pi}\int_y^{\Lambda'} xdx\, \left[ - (\tilde{f}(x) + \pi m) v_I(m,x) + v_+(m,x) \right]F(m,x) \,,
\label{inteq1prime}\\
v_+(m,y) &= i + \frac{1}{2\pi} \int_y^{\Lambda'} xdx\, \left[ x^2(g(x)-1)v_I(m,x) + (\tilde{f}(x)-\pi m) v_+(m,x) \right]F(m,x) \,.
\label{inteq2prime}
\end{align}
Here we have introduced a cutoff $\Lambda' = \beta \Lambda$ ($v_I$ and $v_+$ depend implicitly on $\Lambda'$).
In solving the SD equation for the vertex it is important to keep a finite cutoff, even when the loop integrals converge.
This is due to naive divergences that may change the answer if we remove the cutoff prematurely.
The cutoff should be removed only once the final expression for the correlator is obtained.

To convert \eqref{inteq1prime}, \eqref{inteq2prime} to differential equations, differentiate with respect to $y$.
Everything is now a function of $m,y,\Lambda'$:
\begin{align}
-\frac{2\pi}{yF}\dho_y \left( 
\begin{matrix}
v_I \\
v_+
\end{matrix}
\right) &= \left( 
\begin{matrix}
-\tilde{f} - \pi m & 1\\
y^2(g-1) & \tilde{f} - \pi m 
\end{matrix}
\right) 
\left(
\begin{matrix}
v_I \\
v_+
\end{matrix}
\right) \,.
\label{matrixeqn}
\end{align}
Multiply
the first equation by $(\pi m - \tilde{f}(y))$ and add to the second.
This gives
\begin{align}
  - \frac{2\pi}{yF} \left[ 
  (\pi m - \tilde{f}) \dho_y v_I + \dho_y v_+ \right] =
  -(\pi^2 m^2 + \tily^2) v_I \ed
  \label{foo}
\end{align}
Using \eqref{F} to write $F$ in terms of $f$, \eqref{foo} becomes
\begin{align}
  \dho_y \left[ (\pi m - \tilde{f}) v_I + v_+ \right] &= 0.
\end{align}
We integrate this and obtain
\begin{align}
  (\pi m - \tilde{f}) v_I + v_+ &= \eta(m) \ed
  \label{sol1}
\end{align}
Let us find the integration `constant' $\eta(m)$.
Notice from \eqref{inteq1prime},\eqref{inteq2prime} that $v_I(m,y=\Lambda') = 0$ and $v_+(m,y=\Lambda') = i$.
Using this and substituting $y=\Lambda'$ in \eqref{sol1} gives $\eta(m) = i$.
Plugging $v_+$  from \eqref{sol1} back into the first equation in \eqref{matrixeqn}, we find a differential equation just for $v_I$.
\begin{align}
  \dho_y v_I(m,y) &= \left[ v_I(m,y) - \frac{i}{2\pi m} \right]
  m y F(m,y) 
 \ed \label{vieq}
\end{align}

Integrating \eqref{vieq} while building in the boundary condition $v_I(m,y=\Lambda') = 0$, and using \eqref{sol1} to get $v_+$, we get the exact vertex functions
\begin{align}
v_I(m,y) &= \frac{i}{2\pi m} \left\{ 1  - \exp \left[ -m
\int_y^{\Lambda'} dy'\, y' F(m,y')  
\right]
\right\} \label{visol} \ec \\
  v_+(m,y) &= i + \left[ \tilde{f}(y) - \pi m \right] v_I(m,y) \ed
\end{align}
Recall that $v_I$ and $v_+$ were defined via \eqref{Vv}, \eqref{V++VI+} and 
\eqref{Vdef}.

As a check on the exact result for the vertex, we have computed 
the $O(\lambda)$ vertex correction diagram explicitly and checked that it matches the small $\lambda$ expansion of the exact result.

\subsubsection{Computation of the $T>0$ current two point function}

\begin{figure}[h]
  \centering
  \includegraphics[width=0.4\textwidth]{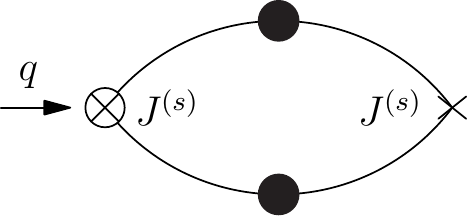}
  \caption{Diagrams contributing to the current 2-point function.
  The notation is the same as in Figure \ref{fig:ferSD}.}
  \label{fig:fermion-JJ}
\end{figure}

With the exact fermion propagator and exact vertex in hand, we are ready to compute the exact current 2-point function $\langle J^+ J^-(q) \rangle$.
It is given by a loop integral shown in Figure \ref{fig:fermion-JJ}.
Notice that we have one exact vertex and one tree-level vertex (using two exact vertices would lead to double counting).
The 2-point function can be written as
\begin{align}
  \langle J^+ J^-(q) \rangle &=
  - \sum_{i=1}^N \int \frac{d^2k}{(2\pi)^2}
  \frac{1}{\beta} \sum_{n=-\infty}^\infty \trace \left[ 
  S_i(k+q) V^+(q,k) S_i(k) V^-_{(0)}
  \right] \ed
\end{align}
The trace here is over the spinor indices, $i$ is the color running in the loop, and $V^-_{(0)} = i \gamma^-$ is the insertion of $J^-$ at tree-level.
The minus sign is due to the fermion loop.
Computing the spinor trace, we have
\begin{align}
  -\frac{N}{\beta} \int &\frac{d^2k}{(2\pi)^2}
  \int_{-1/2}^{1/2} du 
  \sum_{n=-\infty}^\infty
  \frac{1}
  {[\tilde{k}^2 + T^2 \mu_F^2] [(\tilde{k}+q)^2 + T^2 \mu_F^2]} \times
  \cr &\left\{
  2i v_+(m,x) \left[ 
  \tilde{k}_3 (\tilde{k}_3 + q_3) + f(x) k_s (f(x) k_s + i q_3)
  \right] +
  \beta k_s^2 (g(x) - 1)(2f(x) k_s + i q_3) v_I(m,x) \right\}
  \ed \cr 
\end{align}
Using \eqref{sum1},\eqref{sum2} to compute the sums, and \eqref{F} to compute the holonomy integral, we find
\begin{align}
  -\frac{i \beta N}{8\pi\lambda} \int \frac{d^2k}{(2\pi)^2} F(m,x) \cdot
  &\Big\{
  2i (x^2 + \mu_F^2) v_+(m,x) + 
  2i \beta^2 k_s f(x) (f(x) k_s + iq_3) v_+(m,x) + 
  \cr &\quad 
  \beta^3 k_s^2 (g(x) - 1) (2f(x) k_s + iq_3) v_I(m,x)
  \Big\}
  \ed
\end{align}
The angular integral is trivial, because the integrand does not depend on the angle.
We are left with the radial integral,
\begin{align}
  \langle J^+ J^-(m) \rangle =
  \frac{N}{8\pi^2 \lambda\beta}
  \int_0^{\Lambda'} \! dx \, x F(m,x) \cdot &\Big\{
  \left[ x^2 + \mu_F^2 + x^2 f(x)^2 + 2\pi m i x f(x) \right] v_+(m,x)
  + 
  \cr &\quad 
   x^2 (1 - g(x)) (ix f(x) - \pi m) v_I(m,x)
  \Big\} \ed
\end{align}

One can check that this 2-point function has a linear divergence, which we can subtract by adding a counter-term that is proportional to $\Lambda \int \! d^3x \, a_\mu(x) a^\mu(x)$. The same divergence is visible in the zero temperature calculations of \cite{GurAri:2012is}. Subtracting this divergence, and using the relation \eqref{sol1} to write $v_+$ in terms of $v_I$, we obtain the final result for the 2-point function.
\begin{align}
  \langle J^+ J^-(\omega_n) \rangle = 
  -\frac{N}{4\pi}i\omega_n
  -\frac{N \lambda}{8\pi} i\omega_n + 
  \frac{N}{8\pi^2 \lambda \beta}
 & \lim_{\Lambda' \to \infty}
  \int_0^{\Lambda'} \! dx 
  \Bigg\{ -2\pi \lambda + x F(n,x) \times \nonumber \\
  &\Big[ 2(ixf(x) - \pi n) (x^2 + \mu_F^2 + \pi i n x f(x) ) v_I(n,x) + \nonumber \\
  &i \left( x^2  + \mu_F^2 + x^2 f^2(x) + 2\pi i n x f(x) \right) 
  \Big]\Bigg\}
  \ed
  \label{JJfer}
\end{align}
The first two terms come from the anomaly as discussed in Section
\ref{sec:zeroT}. The final term is computed using the various Schwinger-Dyson equations, as we just discussed. The $-2\pi \lambda$ in the integrand subtracts off a linear divergence from the integrand. 
The quantities $\mu_F, f(x), F(n,x)$ and $v_I(n,x)$ are defined in equations \eqref{muF}, \eqref{f}, \eqref{F}, and \eqref{visol} respectivelly.

Equation \eqref{JJfer} is our result for the current-current two-point function in the fermionic theory \eqref{Sfer} at large $N$. It is exact in $\lambda$. In Section \ref{sec:integrals} we have explained how 
the longitudinal and Hall conductivities can be computed from \eqref{JJfer} using equations \eqref{sigmaxx} and \eqref{sigmaxy}, respectively.
For extracting the $\lambda$-even and $\lambda$-odd pieces, 
it is useful to note that $f,F$ and $\mu_F$ are all odd in $\lambda$. In Section \ref{sec:physics} we explored the physics contained in the integral \eqref{JJfer}.

As a first check on this result, one can verify that in the zero temperature limit the above results go over to the $\lambda$-exact current-current two-point function computed in \cite{GurAri:2012is}.
As a second check we consider the weak coupling expansion of this result and compare the leading terms with the abelian results in \cite{PhysRevB.57.7157}. 
Expanding the odd piece, we find
\begin{align}
 \frac{1}{i\omega_m} \langle J^+ J^-(\omega_m) \rangle \Big\vert_{\lambda\text{-odd}} &=  -\frac{N \lambda}{4\pi}  \int_0^\infty dx
\frac{\tanh(x/2)}{x^2 + \pi^2 m^2}
\left\{
2 \log\left[2 \cosh \frac{x}{2}\right] - x^2 \int_{x}^\infty dy \frac{\tanh(y/2)}{y^2 + \pi^2 m^2}
\right\} + O(\lambda^3) \,.
\end{align}
This is in precise agreement with \cite{PhysRevB.57.7157} once we set $N=1$.\footnote{For the matching to work, set $q=1, e=1, h=2\pi$ and $\alpha = \lambda$ in \cite{PhysRevB.57.7157}.}

\subsection{A new test of the $3d$ bosonization duality}
\label{scalars}

In this section we compute the finite temperature conductivity in the critical scalar theory
at leading order in $\lambda_b$. The critical scalar theory can be obtained from the scalar action \eqref{Sscalar} 
by taking $\lambda_4 \to \infty$. We then show that it agrees perfectly with the strong-coupling limit  ($\lambda\to 1$) of the fermionic result obtained above.

While unnecessary for our leading order computation in $\lambda_b$, the scalar propagator
can be computed exactly in $\lambda_b$. This was done in \cite{Aharony:2012ns}, and the result is
\begin{align}
  \langle \phi_i(p) \phi^\dagger_j(-q) \rangle = \delta_{ij}
  \left[ \frac{1}{\tilde{p}^2 + T^2 \mu^2} \right]_{ii} \cdot (2\pi)^3
  \delta(p-q) \,.
\end{align}
Here $\tilde p$ is the shifted momentum (\ref{eq:shift}), and the scalar thermal mass
$\mu$ is given by
\begin{align}
  \mu = \frac{2}{\pi \lambda_b} \text{Im} \, 
  \Li_2 \left( e^{-\mu-\pi i \lambda_b} \right) \ed
  \label{muB}
\end{align}
The exact thermal mass \eqref{muB} can be expanded as
\begin{align}
  \mu(\lambda_b) = \mu_0 + O(\lambda_b^2) \ecq
  \mu_0 = 2 \log \left( \frac{1 + \sqrt{5}}{2} \right) \ed
\end{align}

\begin{figure}[h]
  \centering
  \includegraphics[width=1.0\textwidth]{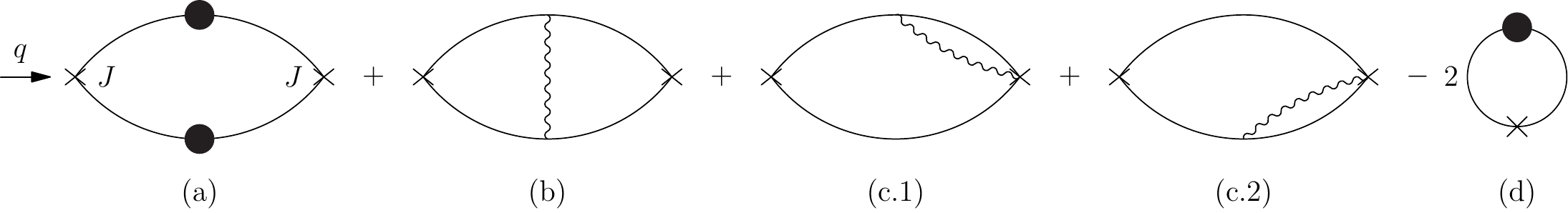}
  \caption{Diagrams contributing to scalar 2-point function up to $O(\lambda_b)$. The notation is as in Figure \ref{fig:ferSD}, with the solid lines now denoting boson propagators.}
  \label{fig:scalar-diagrams}
\end{figure}

The bosonic $U(1)$ current is given in \eqref{Jbos} and the correlator is given by
\begin{align}
  \langle J_\mu J_\nu(p) \rangle 
  - 2\delta_{\mu\nu} \langle \phi^\dagger \phi \rangle
  \ed
\end{align}
The second `diamagnetic' term is due to the seagull diagram that couples the scalars to the background gauge field.
Up to order $\lambda_b$, the diagrams that contribute to the current-current correlator are shown in Figure \ref{fig:scalar-diagrams}. 
The scalar propagators implicitly include the effect of the $(\phi^\dagger \phi)^2$ interaction through the thermal mass $\mu_0$.

\begin{figure}[h]
  \centering
  \includegraphics[width=0.4\textwidth]{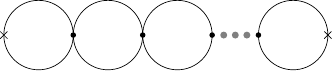}
  \caption{A typical diagram with $\lambda_4$ interaction that is not a purely propagator correction. Diagrams of this type all vanish.}
  \label{fig:lambda4-diagrams}
\end{figure}

Additional diagrams that include the $\lambda_4$ interaction (shown in Figure \ref{fig:lambda4-diagrams}) vanish for the following reason.
The sum of planar diagrams with $n$ $(\phi^\dagger \phi)^2$ vertices and any number of gluons is given by
\begin{align}
  \left( - \frac{\lambda_4}{N} \right)^n
  \langle J_b^+ \cO \rangle 
  \langle \cO \cO \rangle^{n-1}
  \langle \cO J_b^- \rangle \ec
\end{align}
where $\cO = \phi^\dagger \phi$.
The last factor vanishes at zero spatial momentum due to rotational invariance.

The Feynman diagrams (a) and (d) in Figure \ref{fig:scalar-diagrams} are $\lambda$-even and contribute at  $O(\lambda_b^0)$. 
Their sum gives
\begin{align}
  - \frac{N_b\Lambda}{4\pi}
  - \frac{N_b}{4\pi\beta} (\pi^2 n^2 + \mu_0^2)
  \int_{\mu_0}^{\Lambda} dx
  \frac{\coth \left(x/2 \right)}{x^2 + \pi^2 n^2}
  \ed
\end{align}
This expression has a linear divergence that can be subtracted by adding a counter-term proportional to the background $U(1)$ field squared $\Lambda \int a_\mu a^\mu$.
The contributions at $O(\lambda_b)$ come from the graphs (b) and (c) in Figure 
\ref{fig:scalar-diagrams}.
The full renormalized two-point function is then
\begin{align}
  \langle J_b^+ J_b^-(\omega_n) \rangle =
  \frac{N_b\lambda_b}{8\pi} i\omega_n &- \frac{N_b (\pi^2 n^2 + \mu_0^2)}{4\pi\beta}
  \int_{\mu_0}^{\infty} dx
  \frac{\coth(x/2)}{x^2 + \pi^2 n^2} \nonumber \\
  &- \frac{iN_b\lambda_b n (\pi^2 n^2 + \mu_0^2 )}{4\beta}
  \left[ \int_{\mu_0}^{\infty} dx 
  \frac{\coth(x/2)}{x^2 + \pi^2 n^2}
  \right]^2 + O(\lambda_b^2) \ed
  \label{JJsc}
\end{align}
The first term (which does not come from the diagrams shown above) is required for canceling an anomaly, as discussed in Section \ref{sec:zeroT}.

Close to the strong coupling limit $\lambda \to 1$ in the fermion theory, 
one can take the exact fermionic two-point function \eqref{JJfer}, convert the fermionic parameters to bosonic ones via \eqref{map}, and expand the result to order $O(\lambda_b)$. We have omitted the details, which are tedious but straightforward. The end result is that this strong coupling expansion of the fermion agrees precisely with 
(\ref{JJsc}).
(See also Figure~\ref{fig:compare} which compares the results of the two theories in the weak $\lambda_b$ regime.) Again, the pieces of the correlators coming from the anomaly are crucial for the answers to agree.
This constitutes a new test of the bosonization duality at finite temperature. An entire function of $\omega$ has been matched.

\section{Discussion}
\label{sec:discuss}

The main technical result of this paper has been the solution of the Schwinger-Dyson equation for the $T>0$ current-current correlator in fermionic Chern-Simons-matter theory, at $N=\infty$ and exactly in the 't Hooft coupling $\lambda$. These exact results have allowed us to match an entire function of $\omega/T$ between the two sides of the conjectured $3d$ bosonization duality for Chern-Simons-matter theories. As part of this match, we have resolved a discrepancy in the Hall conductivity that is relevant already at $T=0$.

The current-current correlators determine the physical observables $\sigma_{xx}(\omega)$ and $\sigma_{xy}(\omega)$ --- the longitudinal and Hall conductivities. We have explained in Section \ref{sec:physics} how these observables characterize dissipation in these theories as well as the non-dissipative Hall conductivities. The results are qualitatively similar to the Abelian Chern-Simons-matter results previously obtained in \cite{PhysRevB.57.7157}, demonstrating the robustness of the behavior of these functions towards the exact treatment of a certain class of interactions.

We have emphasized in Section \ref{sec:div} that a divergence in the $\sigma_{xx}(\omega)$ conductivity at $\omega = 0$ survives the exact treatment of interactions at $N = \infty$. We explained in Appendix \ref{app:twopt} that this is due to the not-entirely-trivial survival of an infinite tower of `sufficiently conserved' high spin currents in the large $N$ theory that overlap with the electrical current operator. The delicate point here was that there are operators with order $N$ vacuum expectation values at $T>0$. These can ruin the conservation of the high spin currents even at large $N$ inside two point functions. We have shown, however, that quantities that are conserved inside nonzero temperature two point correlation functions can be defined.

The most immediate open question from this work concerns the resolution of the divergence at $\omega = 0$ by finite $N$ effects, ideally still working exactly in $\lambda$. A possible approach to this problem uses the memory matrix formalism. Here the conductivity is written \cite{forster}
\be\label{eq:mem}
\sigma_{ab}(\omega) = \sum_{c,d} \rchi_{J_a Q_c} \left( \frac{1}{- i \omega \rchi + M(\omega) + {\mathcal N}} \right)_{Q_c Q_d} \rchi_{Q_d J_b} \,.
\ee
In this expression the $Q_a$ are the same operators that we considered in Section \ref{sec:div}, describing conserved quantities in the $N=\infty$ theory. The $\rchi$'s quantify the overlap of the $Q_a$ operators with the electric current. An extended recent and explicit discussion of this formalism (including the situation when time reversal invariance is broken, of relevance to Chern-Simons theories) can be found in \cite{Davison:2016hno}. The point is that in the $N = \infty$ theory the divergence in the conductivity at $\omega = 0$ arises because $M(0) = {\mathcal N} = 0$ in (\ref{eq:mem}).
This vanishing occurs because $M(0)$ and ${\mathcal N}$ are proportional to correlation functions of time derivatives of the $Q_a$ operators, and $\dot Q_a = 0$ in the $N=\infty$ theory. At large but finite $N$, we can use the fact that
the $\dot Q_a$ operators are themselves, as operators, proportional to inverse powers of $N$. Therefore, to obtain the leading order answers for $M(0), {\mathcal N} \sim 1/N^\#$, the correlation functions in $M$ and ${\mathcal N}$ may be evaluated in the $N = \infty$ theory. In this way the resolution of the divergence, which is a $1/N$ effect, is reduced to the computation of (infinitely many) correlation functions of conserved currents in the $N = \infty$ theory. Perhaps these computations are feasible. For this approach to give the correct answer, the operators $Q_a$ must saturate the inequality (\ref{eq:dfinal}).

While the delta function is resolved at finite $N$, as we have just described, at order $1/N$ a qualitatively new divergence will appear in the $\omega \to 0$ conductivity. This will be a logarithmic divergence due to a `late-time tail' caused by hydrodynamic charge fluctuations. See e.g. \cite{Kovtun:2003vj}. To our knowledge the resolution, if any exists, of these divergences is not understood. This question might be accessible in vector large $N$ theories, such as the one we have been studying. The challenge will be to identify the correct set of finite $N$ diagrams to resum.

\section*{Acknowledgements} 
The authors would like to thank Ofer Aharony, Maissam Barkeshli, Lorenzo Di Pietro, Blaise Gout\'eraux, Steve Kivelson, Zohar Komargodski, William Witczak-Krempa and Ran Yacoby for stimulating discussions and comments. The work of S.A.H. is partially supported by a DOE Early Career Award.
The work of G.~G. is supported by a grant from the John Templeton Foundation. 
The opinions expressed in this publication are those of the authors and do not necessarily reflect the views of the John Templeton Foundation.
RM  is supported by a Gerhard Casper Stanford Graduate Fellowship.

\appendix

\section{Conventions}
\label{app:conventions}

The conventions used in this work follow those of \cite{Aharony:2012ns}.
The Chern-Simons-matter theories are defined on a spatial plane with coordinates $x^1=x,x^2=y$, and $x^3$ denotes the coordinate along the thermal circle with periodicity $\beta = 1/T$.
The $U(N)$ gauge field is given by $A^\mu = A^\mu_a T^a$ where $T^a$ are anti-Hermitian $U(N)$ generators with the normalization $\trace_N (T^a T^b) = -\frac{1}{2} \delta^{ab}$.
They obey the identity $(T^a)_{ij} (T^a)_{kl} = -\frac{1}{2} \delta_{il} \delta_{jk}$.
In the fermion theory the spinor matrices are given by the Pauli matrices $\gamma^\mu = \sigma^\mu$, $\mu=1,2,3$.

Define the `lightcone' coordinates $x^\pm = (x^1 \pm i x^2)/\sqrt{2}= (x \pm i y)/\sqrt{2}$.
In these coordinates $\delta_{33}= \delta_{+-} = 1$, $\epsilon_{+-3} = i$.
For a 3-momentum $p$ we also define $p_s^2 = 2p^+ p^-$ so that the momentum obeys $p^2 = p_3^2 + p_s^2$.
We will sometimes use the dimensionless variables
\begin{align}
  x = \beta k_s \ecq
  y = \beta p_s \ecq
  \tilde{x} = \sqrt{x^2 + \mu_F^2} \ed
\end{align}
Here $\mu_F$ is the dimensionless thermal mass of the fermion, given in \eqref{muF}.

We work in `lightcone' gauge $A_- = 0$, where the Chern-Simons action \eqref{SCS} becomes
\begin{align}
  \frac{k}{4\pi} \int \! d^3x \, A^a_+ \dho_- A^a_3 \ed
\end{align}
The gauge field propagator is
\begin{center}
  \begin{tabular}{m{3cm}l}
    \includegraphics[width=0.2\textwidth]{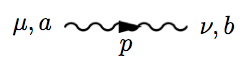}
    & $\displaystyle = G_{\nu\mu}(p) \delta_{ab} \ec$
  \end{tabular}
\end{center}
and its only non-vanishing components are
\begin{align}
  G_{+3}(p) = -G_{3+}(p) = \frac{4\pi i}{k} \frac{1}{p^+} \ed
\end{align}
To regulate the momentum integrals we use a hard cutoff $\Lambda$ in the $x\! -\! y$ plane, and define the dimensionless cutoff $\Lambda' = \beta \Lambda$.

\section{Time-averaged 2-point functions and conserved quantities}
\label{app:twopt}

In this section we prove a relation between the long-time behavior of 2-point functions and constants of motion in large $N$ field theories.
We then apply this relation to Chern-Simons matter theories.
The argument presented here is an adaptation of an argument by Suzuki \cite{suzuki}.
It should be intuitively plausible that late time dynamics is controlled by conserved quantities.

Let $A(t)$ be a single-trace operator at zero spatial momentum, and consider the time-averaged finite temperature correlator
\begin{align}
  C_{AA} = \lim_{t_o \to \infty} \frac{1}{t_o} \int_0^{t_o}
  \langle A^\dagger(0) A(t) \rangle dt \ed
\end{align}
Let us first show that
\begin{align}
  C_{AA} \ge 0 \ed
  \label{CApos}
\end{align}
Expanding in a basis of energy eigenstates,
\begin{align}
  C_{AA} &=
  \lim_{t_o \to \infty} 
  \frac{1}{t_o} \int_0^{t_o} 
  \sum_{k,l} e^{-\beta E_k} e^{i(E_l-E_k)t}
  \bra{k} \! A^\dagger \! \ket{l} 
  \bra{l} \! A \! \ket{k}
  \cr &=
  \sum_{
  \begin{smallmatrix}
  k,l \\
  E_k=E_l
  \end{smallmatrix}
  } e^{-\beta E_k} 
  \left| \bra{l} \! A \! \ket{k} \right|^2
  \ge 0 \ed
  \label{eq-timeaveragespectral}
\end{align}
When $A=J$, this is the Drude weight as written in equation (\ref{eq-drudespectral}).

Now, let $Q_a$ be a set of single-trace operators that are constants of motion within any 2-point function $\langle Q_a \cO \rangle$ where $\cO$ is a single-trace operator.
In particular, we assume that
\begin{align}
  \langle \dot{Q}_a(0) \cO(t) \rangle = 0
  \label{Qdot}
\end{align}
up to $O(1/N)$ corrections.

We can then write $A = \alpha_a Q_a + A'$ where $A'$ is orthogonal to all $Q_a$, and $\alpha_a = \langle Q^\dagger_a A \rangle / \langle Q^\dagger_a Q_a \rangle$.
(Orthogonality is defined with respect to the inner product $(A,B) = \langle A^\dagger B \rangle$, where $A,B$ are two operators at zero momentum and at equal times.)
The time-averaged correlator is now given by
\begin{align}
  C_{AA} &= 
  \lim_{t_o \to \infty} 
  \frac{1}{t_o} \int_0^{t_o} 
  \left[ 
  \alpha_a^* \alpha_b \langle Q_a^\dagger Q_b(t) \rangle
  + \alpha_a^* \langle Q_a^\dagger A'(t) \rangle
  + \alpha_b \langle A'^\dagger Q_b(t) \rangle
  + \langle A'^\dagger A'(t) \rangle
  \right] 
  \cr &=
  \alpha_a^* \alpha_b \langle Q_a^\dagger Q_b \rangle
  + \alpha_a^* \langle Q_a^\dagger A' \rangle
  + \alpha_b \langle A'^\dagger Q_b \rangle
  + \lim_{t_o \to \infty} 
  \frac{1}{t_o} \int_0^{t_o} 
  \langle A'^\dagger A'(t) \rangle
  \cr &=
  \alpha_a^* \alpha_b \langle Q_a^\dagger Q_b \rangle
  + \lim_{t_o \to \infty} 
  \frac{1}{t_o} \int_0^{t_o} 
  \langle A'^\dagger A'(t) \rangle \ed
\end{align}
In the second equality we used the fact that $Q_a$ is constant inside 2-point functions, and in the last equality we used the orthogonality of $Q_a$ with $A'$.
In the resulting expression, the last term is non-negative by the same argument that led to \eqref{CApos}.
We therefore arrive at the conclusion,
\begin{align}
  C_{AA} \ge 
  \sum_{a,b} \frac{
  \langle A^\dagger Q_a \rangle 
  \langle Q_a^\dagger Q_b \rangle
  \langle Q_b^\dagger A \rangle
  }
  {
  \langle Q_a^\dagger Q_a \rangle
  \langle Q_b^\dagger Q_b \rangle
  } 
  \ed \label{CAineq}
\end{align}
Notice that the right-hand side is non-negative because it involves the positive-definite Hermitian matrix $\langle Q_a^\dagger Q_b \rangle$.
The equality in \eqref{CAineq} holds if
\begin{align}
  \lim_{t_o \to \infty} 
  \frac{1}{t_o} \int_0^{t_o} 
  \langle A'^\dagger A'(t) \rangle = 0 \ed
\end{align}

\subsection{Free fermion}

Let us first discuss the inequality \eqref{CAineq} in the context of the free fermion theory.
We will focus on the 2-point function of the electric current, taking $A = J_x$, and show that the long-time correlator $C_{J_x J_x}$ is positive.
As discussed in Section~\ref{sec:div}, this implies that the longitudinal conductivity will have a term of the form $D \delta(\omega)$ with $D$ positive.

The free theory has infinitely many conserved high-spin currents $J^{(s)}_{\mu_1\cdots \mu_s}$ with spins $s=1,2,3,\dots$.
They are given by the generating function \cite{Giombi:2011kc}
\begin{align}
  \cO(x,\epsilon) = \sum J^{(s)}_{\mu_1 \cdots \mu_s}
  \epsilon^{\mu_1} \cdots \epsilon^{\mu_s} =
  \bar{\psi} 
  (\vec{\gamma} \cdot \vec{\epsilon}) \, 
  f(\overrightarrow{\dho_\mu},\overleftarrow{\dho_\mu},\vec{\epsilon})
  \psi \ec \label{Jgen}
\end{align}
where
\begin{align}
  f(\vec{u}, \vec{v}, \vec{\epsilon}) = \frac{
  \exp \left( \vec{u}\cdot\vec{\epsilon} - \vec{v}\cdot\vec{\epsilon} \right)
  \sinh \sqrt{
  2 \vec{u}\cdot\vec{v} \, \vec{\epsilon}\cdot\vec{\epsilon}
  - 4 \vec{u} \cdot \vec{\epsilon} \, \vec{v} \cdot \vec{\epsilon}
  } }
  { \sqrt{
  2 \vec{u}\cdot\vec{v} \, \vec{\epsilon}\cdot\vec{\epsilon}
  - 4 \vec{u} \cdot \vec{\epsilon} \, \vec{v} \cdot \vec{\epsilon}
  } } \ed
\end{align}
Notice that these conserved currents are completely symmetric by construction.

As before, we will denote the spin 1 current by $J = J^{(1)}$.
From each current $J^{(s)}$ with $s \ge 2$ we can build a conserved vector charge
\be
Q^{(s)}_i = \int d^2x J^{(s)}_{tt\cdots ti}(x) \,,
\label{Q}
\ee
because
\begin{align}
  \frac{d}{d t} Q^{(s)}_i =
  \frac{d}{d t} \int \! d^2x \, J^{(s)}_{tt\cdots ti}(x) =
  - \int \! d^2x \, \dho_j J^{(s)}_{jt\cdots ti}(x) = 0 \ed
\end{align}
Here we dropped a surface term.
From \eqref{CAineq}, these conserved vector charges will contribute to the long-time correlator $C_{J_x J_x}$ (and hence to the Drude weight in the conductivity) if they overlap with the electric current, \textit{i.e.} if $\langle J_x Q^{(s)}_x \rangle \ne 0$ at finite temperature. Two vector operators will generally overlap if allowed by symmetries. The relevant symmetry here is charge conjugation. The current $J_x$ is odd under charge conjugation. 
As we now show, odd spin currents are also odd under charge conjugation, whereas the even spin charges are even. Therefore, we expect the operators $Q^{(3)}, Q^{(5)}, \dots$ to overlap with the current operator.

To see this we work in Euclidean space, and recall that in our conventions $\bar{\psi} = \psi^\dagger$.
Charge conjugation can be defined by
\begin{align}
  \psi \to i\sigma^2 \psi^* \ecq \psi^\dagger \to i \psi^T \sigma^2 \ed
\end{align}
It is easy to check that under charge conjugation the generating function \eqref{Jgen} transforms as $\cO(x,\epsilon) \to \cO(x,-\epsilon)$.
Therefore, even spin currents are even under charge conjugation, while odd spin currents are odd, and the electric current can overlap with all currents $J^{(s)}$ with spins $s=3,5,\dots$.

\subsection{Chern-Simons matter theories}

The situation for Chern-Simons matter theories is more complicated because the high-spin currents in these theories are not exactly conserved.
However, we will now show that in the Chern-Simons fermion theory one can still construct approximate constants of motion that contribute to the time-averaged correlator $C_{J_x J_x}$.
This implies that the Drude weight is positive also in Chern-Simons matter theories at large $N$.

The spectrum of primary single-trace operators of the Chern-Simons fermion theory consists of current operators $J^{(s)}$ with spins $s=1,2,3,\dots$, and of a scalar operator $J^{(0)} = \bar{\psi} \psi$.
The electric current $J=J^{(1)}$ and the stress tensor $J^{(2)}$ are conserved exactly, while the higher spin currents are conserved only up to multi-trace terms \cite{Giombi:2011kc}.
Schematically, for $s>2$ we have
\begin{align}
  \dho \cdot J^{(s)} = \frac{f_1(\lambda)}{N} J J  +
  \frac{f_2(\lambda)}{N^2} J J J  \ec
\end{align}
where the terms on the right are double-trace and triple-trace operators (here $J$ denotes a general single-trace primary), with possible additional derivatives.
As a result, the constants of motion \eqref{Q} that we construct from these currents are conserved in time up to multi-trace operators.

In order to derive the inequality \eqref{CAineq} for the time-averaged correlator, we must show that these constants of motion are conserved inside 2-point functions as in equation \eqref{Qdot}.
This is true at zero temperature, where multi-trace operators do not overlap with single-trace operators at large $N$.
(This implies, for example, that the operators $J^{(s)}$ do not acquire an anomalous dimension at large $N$ for any spin.)
But at nonzero temperature this is no longer true in general, because single-trace operators can have non-zero (and order $N$) expectation values.
Indeed, in this case we can have (schematically) $\langle \dho \cdot J^{(s)} \cO \rangle \sim \frac{1}{N} \langle JJ \, \cO \rangle \sim \frac{1}{N} \langle J \rangle \langle J \cO \rangle$, and this will contribute to the non-conservation of $Q_{(s)}$ at leading order at large $N$.

Let us focus on the spin 3 case for concreteness.
In this case we will see that there is a single term that contributes to the non-conservation of $Q_{(3)}$ at leading order.
Further, we will show that there is an improvement of the current that allows us to remove this term, leading to a constant of motion that satisfies the conservation equation \eqref{Qdot}.

The operator $J^{(3)}$ has dimension $\Delta_3 = 4 + O(1/N)$.
Let us work out the most general operator form of the divergence $\dho^\mu J^{(3)}_{\mu\nu\rho}$.
It is a primary operator with dimension 5 and spin 2 that is odd under charge conjugation.
There are no single-trace primary operators with these quantum numbers in the spectrum, and so the divergence is equal to a sum of multi-trace operators.
On dimensional grounds, the only single-trace operators that can participate in these multi-trace operators are $J^{(0)}$, $J^{(1)}$, and $J^{(2)}$, as well as possible derivatives.
All the multi-trace operators that can appear in the divergence of $J^{(3)}$ have the schematic form $\dho J^{(0)} J^{(1)}$.
Indeed, operators of the form $\dho J^{(0)} J^{(0)}$, $\dho J^{(1)} J^{(1)}$, and $J^{(2)} J^{(0)}$ are ruled out because they are even under charge conjugation. $J^{(2)} J^{(1)}$ is ruled out because of the triangle inequality.\footnote{
The triangle inequality is the following statement about zero temperature correlators: In a 3-point function of the form $\langle J^{(s_1)} J^{(s_2)} J^{(s_3)} \rangle$ where $s_1 \le s_2 + s_3$, the current $J^{(s_1)}$ is conserved \cite{Maldacena:2012sf}.
Therefore, a $J^{(2)} J^{(1)}$ term cannot appear in $\dho \cdot J_3$ because then $J^{(3)}$ would not be conserved inside $\langle J^{(3)} J^{(1)} J^{(2)} \rangle$.
}
Finally, triple-trace operators are ruled out because of dimensions.

We find that $\dho \cdot J^{(3)} \sim \frac{1}{N} \dho J^{(0)} J^{(1)}$, where the combination on the right-hand side is schematic and can appear with different sprinklings of the indices.
The allowed combinations are
\begin{align}
  \dho^\rho J^{(3)}_{\rho \mu \nu} = 
  \frac{a_1}{N} \eta_{\mu\nu} \dho_\alpha J^{(0)} \cdot J^{(1)}_\alpha
  + \frac{a_2}{N} \dho_{(\mu} J^{(0)} \cdot J^{(1)}_{\nu)}
  + \frac{a_3}{N} J^{(0)} \cdot \dho_{(\mu} J^{(1)}_{\nu)} 
  \ec \label{J3div}
\end{align}
where the coefficients $a_i$ depend on $\lambda$.
Let us see what this implies for the conservation equation \eqref{Qdot}.
Here, the approximate constant of motion is
\begin{align}
  Q_i^{(3)}(t) = \int \! d^2x \, J^{(3)}_{tti}(x,t) \ed
\end{align}
In our case, the other insertion in equation \eqref{Qdot} is a vector under rotations which we denote $\cO_i$. Hence,
\begin{align}
  \langle \dot{Q}_i^{(3)} \cO_i(t) \rangle &= 
  \int \! d^2x \, \langle (\dho \cdot J^{(3)})_{ti}(x,0) \cO_i(t) \rangle 
  \cr &=
  \frac{1}{N} \int \! d^2x \, 
  \Big\langle 
  \left[
  a_2 \dho_{(t} J^{(0)} \cdot J^{(1)}_{i)}
  + a_3 J^{(0)} \cdot \dho_{(t} J^{(1)}_{i)}
  \right]\!(x,0) \,\,
  \cO_i(t) \Big\rangle 
  \ed
\end{align}
The only way to get a leading contribution at large $N$ is if the 3-point function factorizes, and the only factorization allowed by rotation symmetry is $\langle J^{(0)}(x) \rangle \cdot \langle J^{(1)}_i(x) \cO_i(t) \rangle$.
The 1-point function $\langle J^{(0)} \rangle$ is independent of $x$ and $t$ so we can drop the $a_2$ term.
We are left with
\begin{align}
  \langle \dot{Q}^{(3)}_i \cO_i(t) \rangle &= 
  \frac{a_3}{N} \langle J^{(0)} \rangle 
  \int \! d^2x \, 
  \langle \dho_t J^{(1)}_{i}(x,0) \cO_i(t) \rangle 
  \ed \label{QdotO}
\end{align}
The expectation value in this term will generally be of order $N$ and hence this term will spoil the conservation.

To solve this problem, note that we have the freedom to define a new current
\begin{align}
  \tilde{J}^{(3)}_{\mu\nu\rho} =
  J^{(3)}_{\mu\nu\rho} + 
  \frac{a_4(\lambda)}{N} \eta_{(\mu\nu} J^{(1)}_{\rho)} J^{(0)} \ed
  \label{Jt}
\end{align}
Due to parity, $a_4(\lambda)$ is odd in $\lambda$, so this term does not affect the current in the free theory.
The improved constant of motion is given by (taking $\eta_{tt} = 1$)
\begin{align}
  \tilde{Q}^{(3)}_i(t) = Q^{(3)}_i(t) + 
  \frac{a_4(\lambda)}{N} \int \! d^2x \, J^{(0)} J^{(1)}_i(x,t)
  \ed
\end{align}
Then from \eqref{QdotO} we see that
\begin{align}
  \langle \dot{\tilde{Q}}^{(3)}_i \cO_i(t) \rangle &= 
  \frac{a_3 + a_4}{N} \langle J^{(0)} \rangle \int \! d^2x \,
  \langle \dho_t J^{(1)}_{i}(x,0) \cO_i(t) \rangle 
  + O(N^0) \ed
  \label{QtDot}
\end{align}
The leading term can be canceled by choosing $a_4 = -a_3$ , and the redefined charge obeys the conservation equation \eqref{Qdot} as required.

We reached an interesting conclusion.
At zero temperature there is a natural choice for the high spin currents, where each current is the primary operator of a conformal representation with spin $s$.
Technically this means that these currents are symmetric and traceless, which fixes improvement terms such as \eqref{Jt}.
At finite temperature this is no longer a natural choice because conformal symmetry is broken.
The calculation \eqref{QtDot} shows that there is another unique choice for these trace terms that gives the requisite conservation at finite temperature.

\section{Massive fermion at zero temperature}
\label{app:zeroTmassanomaly}

In this section we compute the current-current correlator at $T=0$ with a nonzero fermion mass $\sigma$, that is then taken to infinity.
This computation is used in Section \ref{sec:zeroT} to determine the anomaly in the background $U(1)$ symmetry.

The exact propagator at $T=0$ with fermion mass $\sigma$ is
\begin{align}
  \langle \psi_i(p) \bar{\psi}_j(q) \rangle = S(p) \delta_{ij}
  (2\pi)^3 \delta(p+q) \,,
\end{align}  
where\footnote{
The exact fermion propagator was computed in \cite{Giombi:2011kc}.
Here we compute it in our scheme with a hard cutoff regulator, carefully keeping track of the cutoff dependence.
}
\begin{align}  
  S(p) &= \frac{-i p_\mu \gamma^\mu + i g(y) p_+ \gamma^+ - f(y) p_s I}
  {p^2 + (\sigma \mu)^2} \,, \\
  y f(y) &= (\lambda-\sign(\sigma)) \mu - \lambda \sqrt{y^2 + \mu^2} \,, \\
  f^2(y) - g(y) &= \frac{\mu^2}{y^2} \,, \\
  (\lambda-\sign(\sigma)) \mu + \sign(\sigma) &= 
  \frac{\lambda\mu^2}{2} \frac{1}{\Lambda'} + O(\Lambda^{-3}) \ed
\end{align}
Here
\begin{align}
  x = \frac{k_s}{|\sigma|} \ecq
  y = \frac{p_s}{|\sigma|} \ecq
  z = \frac{q}{|\sigma|} \ecq
  \Lambda' = \frac{\Lambda}{|\sigma|} \ec
\end{align}
$\Lambda$ is the cutoff and $\sigma \mu$ is the renormalized mass.

The exact $J^+$ vertex can be computed using techniques similar to those used in \cite{GurAri:2012is}. 
It is given by
\begin{align}
  \langle J^+(-q) \psi_i(k) \bar{\psi}^j(-p) \rangle =
  V^+(q,p) \delta^i_j (2\pi)^3 \delta(p+q-k) \,,
 \end{align}
 where
 \begin{align}
  V^+(q,p) &= v_+(\bary) \gamma^+ + \frac{2p^+}{q} v_I(\bary) I \,,
  \\
  v_+(\bary) &= 
  \frac{i}{2} \left( 1 + 2i \bar{f}(\bary) \right) 
  - \frac{i}{2} \left( 2i \bar{f}(\bary) - 1 \right)
  e^{2i\lambda \left[ \arctan(2\barL) - \arctan(2\bary) \right]} \,,
  \\
  v_I(\bary) &= 
  \frac{i}{2} - \frac{i}{2}
  e^{2i\lambda \left[ \arctan(2\barL) - \arctan(2\bary) \right]}
  \ed
\end{align}
Here
\begin{gather}
  \barx = \frac{\sqrt{x^2 + \mu^2}}{z} 
  \ecq
  \bary = \frac{\sqrt{y^2 + \mu^2}}{z} 
  \ecq
  \barL = \frac{\sqrt{\Lambda'^2 + \mu^2}}{z}
  \ecq
  \barmu = \frac{\mu}{z}
  \ecq
  \bar{f}(\bary) = \frac{y f(y)}{z}
  \ed
\end{gather}
The 2-point function is given by
\begin{align}
  \langle J^+(-q) J^- \rangle = - N \int \frac{d^3k}{(2\pi)^3}
  \trace \left( S(k) i \gamma^- S(k+q) V^+(k,q) \right) \ed
\end{align}
The bare result is
\begin{align}
  \langle J^+(-q) J^- \rangle =
  - \frac{i N q}{16\pi\lambda} \left\{
  (i-2\barmu)^2 \left[ 
  e^{2i\lambda ( \arctan(2\barL) - \arctan(2\barmu) )} - 1
  \right]
  + 4i \lambda (\barL - \barmu)
  \right\} \ed
\end{align}
This agrees with the known result when $\sigma = 0$ \cite{GurAri:2012is}.
After we remove the linear divergence and take the cutoff to infinity we are left with the following result for the continuum theory:
\begin{align}
  \langle J^+(-q) J^- \rangle =
  \frac{i N q}{16\pi\lambda} \left\{
  \left( 1+\frac{2i \mu\sigma}{q} \right)^2 \left[ 
  \exp \left( \pi i \lambda -
  2 i \lambda \arctan\left( \frac{2\mu|\sigma|}{q} \right) 
  \right) - 1
  \right]
  + \frac{4i \lambda \mu|\sigma|}{q} \right\}
  \ed
\end{align}
In the large mass limit $\sigma/q \to \pm \infty$ we find
\begin{align}
  \langle J^+(-q) J^- \rangle \to
  - \sign(\sigma) \frac{i N q}{4\pi} 
  + \frac{i N \lambda q}{8\pi} 
  + O(1/\sigma) \ed
\end{align}

\section{Plots of fermion conductivities}
\label{app:allplots}

In this appendix we present plots of the real and imaginary parts of the conductivities in the Chern-Simons theory with fermion matter. Various combinations of these quantities have been discussed in section \ref{sec:physics}.

\begin{figure}[H]
\begin{center}
\includegraphics[width=0.8\textwidth]{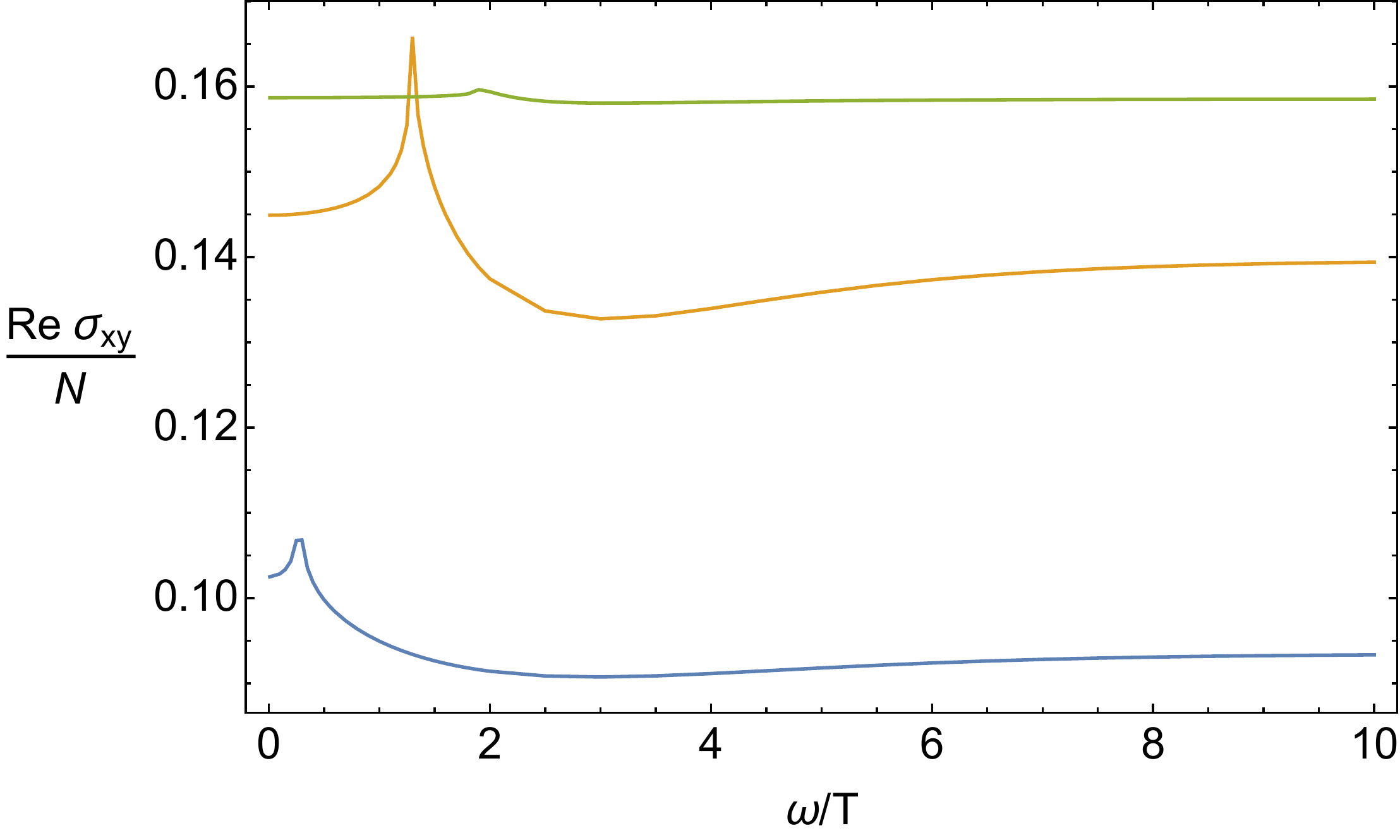}
\includegraphics[width=0.8\textwidth]{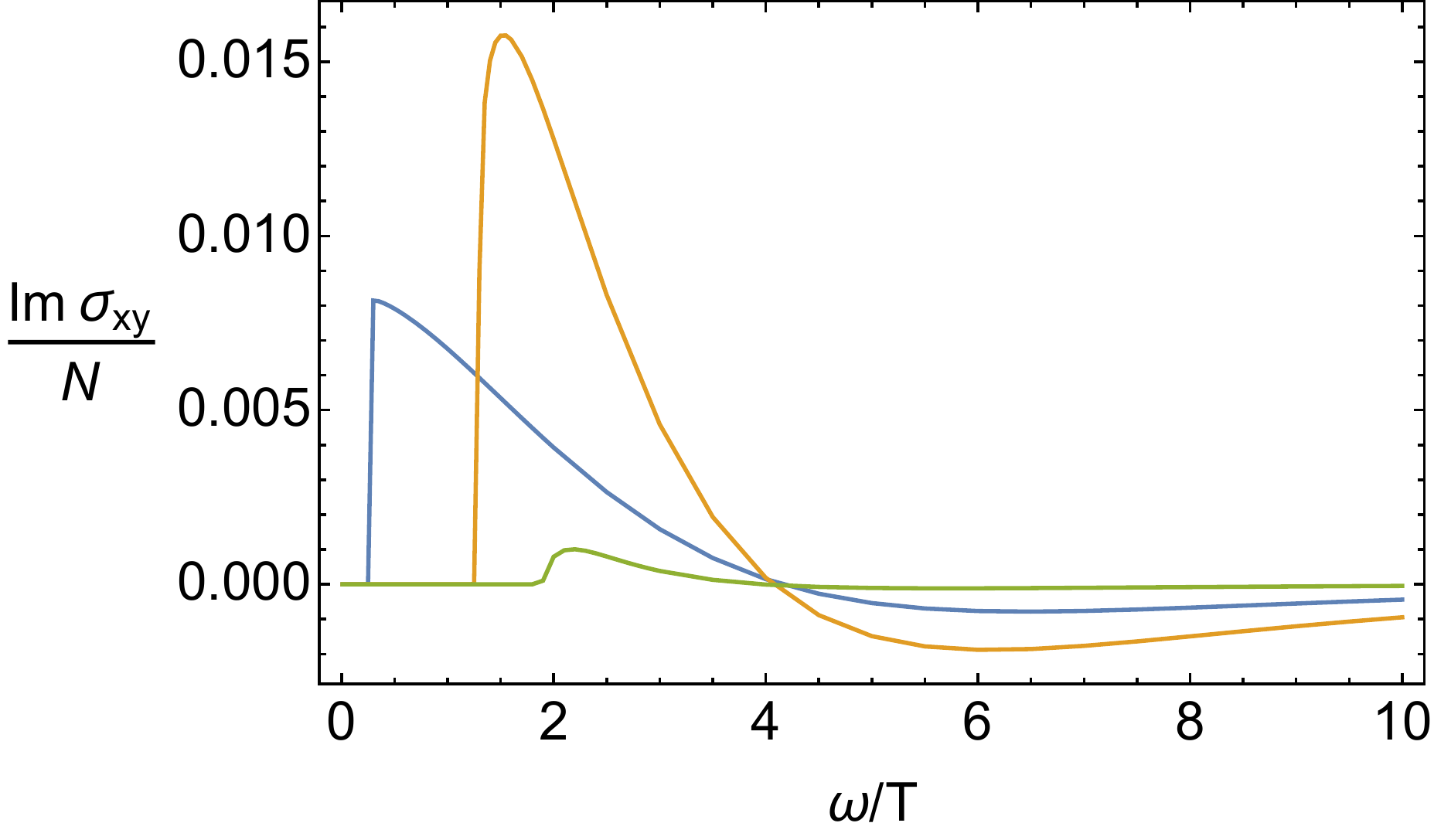}
\end{center}
\caption{The real and imaginary parts of the Hall conductivity $\sigma_{xy}$ as a function of $\omega/T$ for $\lambda$ = 0.1 (blue), 0.5 (orange) and 0.9 (green).}
\end{figure}

\begin{figure}[H]
\begin{center}
\includegraphics[width=0.8\textwidth]{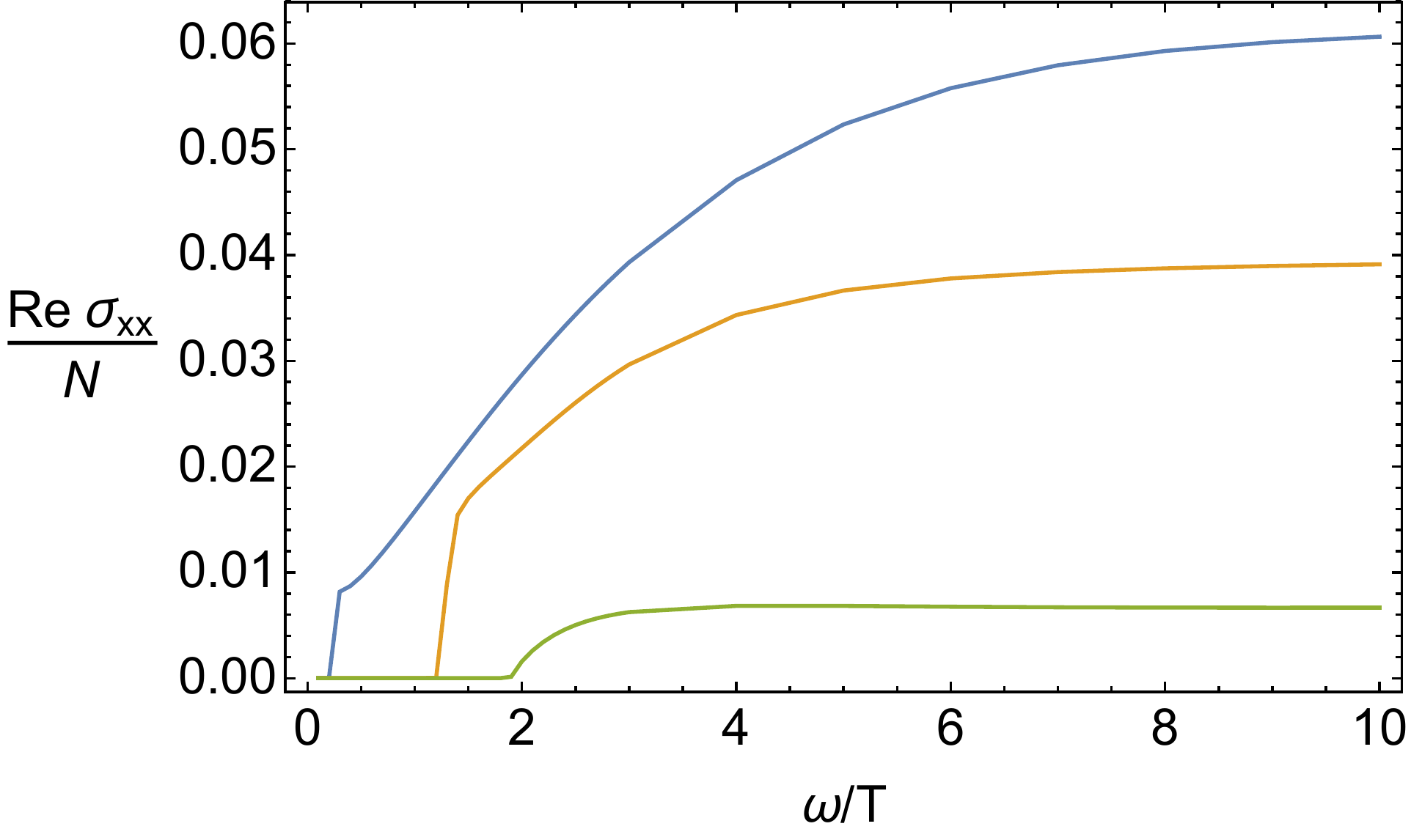}
\includegraphics[width=0.8\textwidth]{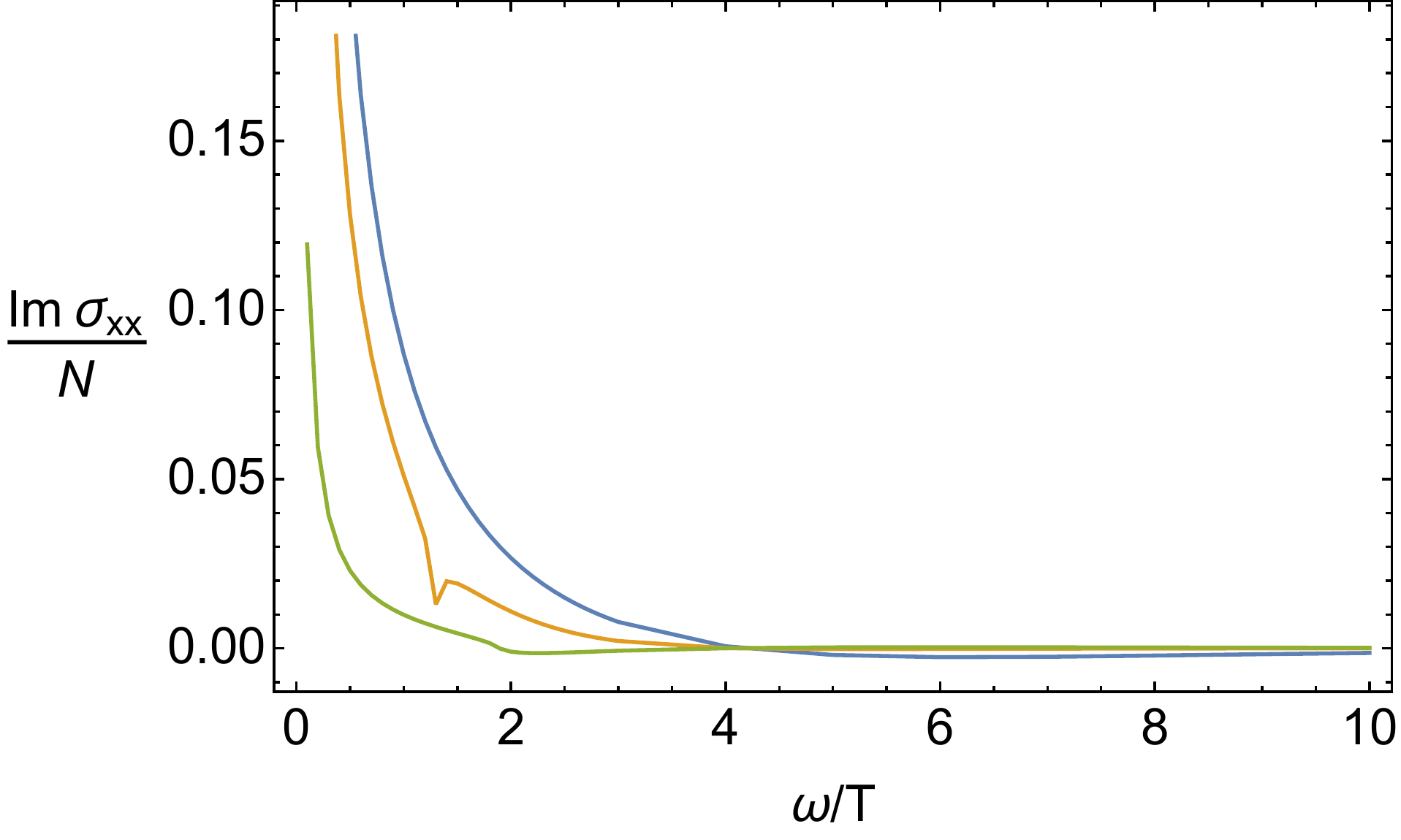}
\end{center}
\caption{The real and imaginary parts of the longitudinal conductivity as a function of $\omega/T$ for $\lambda$ = 0.1 (blue), 0.5 (orange) and 0.9 (green)}.
\end{figure}

\end{document}